\documentclass[preprint,final]{elsarticle}
\usepackage{amsmath}
\usepackage{amsfonts}
\usepackage{lineno,hyperref}
\usepackage[dvipsnames]{xcolor}
\usepackage{subcaption}
\usepackage{xcolor}
\usepackage[utf8x]{inputenc}
\usepackage{chngcntr}
\usepackage{enumitem}
\counterwithout{figure}{subsection}
\usepackage{changes}
\definechangesauthor[name={Ming Fang}, color=orange]{MF}
\usepackage{mathtools}
\usepackage{siunitx}
\usepackage[english]{babel}

\defineshorthand[english]{"~}{\babelhyphen{nobreak}}
\addto\extrasenglish{
  \languageshorthands{english}
  \useshorthands{"}
}
\DeclareSIUnit\Ci{\text{Ci}}
\DeclareSIUnit\Roentgen{\text{R}}
\DeclareSIUnit\gwdpert{\text{GWd/t}}
\DeclareSIUnit\gammas{$\mathrm{\gamma}$}

\newcommand{\software}[1]{\textit{#1}}

\journal{Nuclear Inst. and Methods in Physics Research, A}

\makeatletter
\def\ps@pprintTitle{% \let\@oddhead\@empty \let\@evenhead\@empty
 \def\@oddfoot{}%
 \let\@evenfoot\@oddfoot}
\makeatother

\begin{document}

% \includepdf[pages = 1]{coverletter.pdf}

\begin{frontmatter}

\title{Feasibility of Neutron Coincidence Counting for Spent TRISO Fuel}

% % Group authors per affiliation: 
% \author{M. Fang, N. Bartholomew,and A. Di Fulvio\fnref{myfootnote}} 
% \address{Department of Nuclear, Plasma, and Radiological Engineering, \\University of Illinois, Urbana-Champaign}
% \fntext[myfootnote]{Since 1880.}

% or include affiliations in footnotes:
\author[uiuc]{Ming Fang}
\affiliation[uiuc]{organization={Department of Nuclear, Plasma, and Radiological
                        Engineering, University of Illinois Urbana-Champaign},
                   addressline={104 South Wright Street},
                   city={Urbana},
                   postcode={61801},
                   state={IL},
                   country={United States}}
\author[uiuc]{Angela Di Fulvio\corref{mycorrespondingauthor}}
\cortext[mycorrespondingauthor]{Corresponding author. Tel.: +1 217 300 3769. Fax.: +1 217 333 2906}
\ead{difulvio@illinois.edu}

\begin{abstract}
   High-temperature gas reactors rely on TRIstructural-ISOtropic (TRISO) fuel for enhanced fission product retention. Accurate fuel characterization would improve monitoring of efficient fuel usage and accountability. We developed a new neutron multiplicity counter (NMC) based on boron coated straw (BCS) detectors and used it in coincidence mode for ${}^{235}$U assay in TRISO fuel. In this work, we demonstrate that a high-efficiency version of the NMC encompassing 396 straws is able to estimate the ${}^{235}$U in used TRISO-fueled pebbles or compacts with a relative uncertainty below 2.5\% in \SI{100}{\s}. We performed neutronics and fuel depletion calculation of the HTR-10 pebble bed reactor to estimate the neutron and gamma-ray source strengths of used TRISO-fueled pebbles with burnup between 9 and \SI{90}{\gwdpert}. Then, we measured a gamma-ray intrinsic efficiency of $10^{-12}$ at an exposure rate of \SI[per-mode = symbol]{340.87}{\Roentgen\per\hour}. The low gamma-ray sensitivity and high neutron detection efficiency enable the inspection of used fuel.
%    \end{linenumbers}

\end{abstract}


\end{frontmatter}

% \linenumbers

%------------------------------------------------------------------------------
\section{Introduction}
High temperature gas reactors (HTGRs) are advanced nuclear reactors that have the potential to improve the safety, efficiency, and economics of nuclear energy production~\cite{MOORE1982153,international2010iaea,JASZCZUR20167861}. They rely on tristructural-isotropic (TRISO) fuel, which provides enhanced fission product retention and improved spent fuel management compared to traditional fuels~\cite{DEMKOWICZ2019434,KANIA2013545}. Pebble bed reactors (PBRs) are variants of HTGR, where hundreds of thousands of TRISO-fueled pebbles continuously flow through the reactor core of a PBR and are reinserted until they reach a targeted burnup~\cite{kadak2005future}; the characterization and identification of each individual fuel pebble has the potential to advance both operational and fundamental research aspects of PBRs. For instance, precise determination of fuel transit time can be leveraged to validate computational models, and fuel-use efficiency can be improved by controlling excessive burnup accumulation or premature fuel discharge. Furthermore, enhanced fuel accountability can be achieved by uniquely identifying individual fuel pebbles, which can supplement currently implemented methods for material control and accountability purposes. One of the unique signatures for fuel identification is the \textsuperscript{235}U mass and burnup level, which can be extracted through active interrogation.\replaced{ Passive measurement of \textsuperscript{235}U is not feasible in this scenario due to the low spontaneous fission (SF) rate of \textsuperscript{235}U and limited inspection time. Active neutron interrogation techniques have been extensively used to detect the presence of special nuclear materials~\cite{whetstone2014review,chichester2008using,myers2005photon} and quantify uranium content in nuclear fuel~\cite{gozani1981active,melton2000calibration,eccleston1979measurement}. In active interrogation, a radioactive isotopic source or a generator source is employed to irradiate and induce fission in the sample to be assayed. The fission signatures we focus on in this work are the time-correlated neutron counts emitted by the fuel, which depend on the amount of fissile material and can be measured using the neutron coincidence counting technique~\cite{reilly1991passive,DIFULVIO201792}.}{ Neutron coincidence and multiplicity counting is a non-destructive assay technique that determines the number of time-correlated neutrons emitted by a fissile material to estimate the amount of fissile material present in a sample. The time-correlated neutrons come from spontaneous fission (passive interrogation) or induced fission by irradiating the sample with an external neutron source (active interrogation).} In our previous work~\cite{FANG2023109794}, we developed a high-efficiency neutron multiplicity counter (NMC) based on pie-shape boron coated straw (BCS) detectors to measure the time-correlated neutron counts. We simulated the active interrogation of fresh TRISO-fueled pebbles and estimated the \textsuperscript{235}U mass based on time-correlated counts with a relative uncertainty and error below 2\% in \SI{100}{\s}. 

PBR core typically employs the MEDUL fuel management scheme to achieve higher average fuel burnup and power output, where used fuel pebbles that have not reached the target burnup level are re-inserted into the reactor core~\cite{topan2016study}. Before their re-insertions, measurement of the remaining \textsuperscript{235}U mass in the pebble needs to be performed. However, interrogation of used TRISO-fueled pebbles presents additional challenges compared to fresh pebbles. First, used fuel pebbles are strong passive gamma-ray emitters with gamma-ray intensity up to 10\textsuperscript{13} \replaced{gammas/s/pebble}{cps}~\cite{su2004design}. Therefore, a high-gamma-ray-insensitivity of the BCS-based NMC is crucial to ensure that the gamma-ray background does not interfere with neutron coincidence counting. Second, used fuel pebbles also emit \replaced{($\alpha$,n)}{($\alpha$, n)} neutrons, delayed neutrons\added{ (through decay of delayed neutron precursors)}, and spontaneous fission neutrons\deleted{ through {($\alpha$, n)} reactions, decay of delayed neutron precursors, and spontaneous fission of fission products, respectively}. In particular, spontaneous fission nuclides such as \textsuperscript{240}Pu, \textsuperscript{242}Cm, and \textsuperscript{244}Cm have a high neutron fission yield of $1.02\times10^3$, $2.10\times10^7$, $1.08\times10^7$~n/s/g, respectively~\cite{reilly1991passive}, and their contributions to neutron coincidence counts may become considerable at high burnup levels. In this work, we simulated the neutron interrogation of used fuel pebbles with burnup in the 9-\SI{90}{\gwdpert} range and quantified the contributions to the counts by the gamma-ray background, delayed and \replaced{($\alpha$,n)}{($\alpha$, n)} neutrons, spontaneous fission neutrons, and active interrogation neutrons. We demonstrated the capability of our NMC to accurately estimate the \textsuperscript{235}U mass in a used TRISO-fueled pebble in a short time window of \SI{100}{\s}.

This paper is organized as follows. In Section~\ref{sec:scale_simulation}, computation of isotopics of used fuel pebble and determination of the passive neutron and gamma-ray source terms are performed. In Section~\ref{sec:NMC_description}, a detailed description of the BCS-based NMC system is presented and the high gamma-ray-insensitivity of the NMC is demonstrated. In Section~\ref{sec:active_interrogation}, simulation of neutron active interrogation of partially spent fuel pebble using the BCS-based NMC is carried out. Finally, the discussion and conclusions are presented in Section~\ref{sec:conclusion}.
%------------------------------------------------------------------------------
\section{Computational Method}\label{sec:scale_simulation}
\subsection{Introduction}
\begin{figure}[!htbp]
	\centering
	\includegraphics[width=\linewidth]{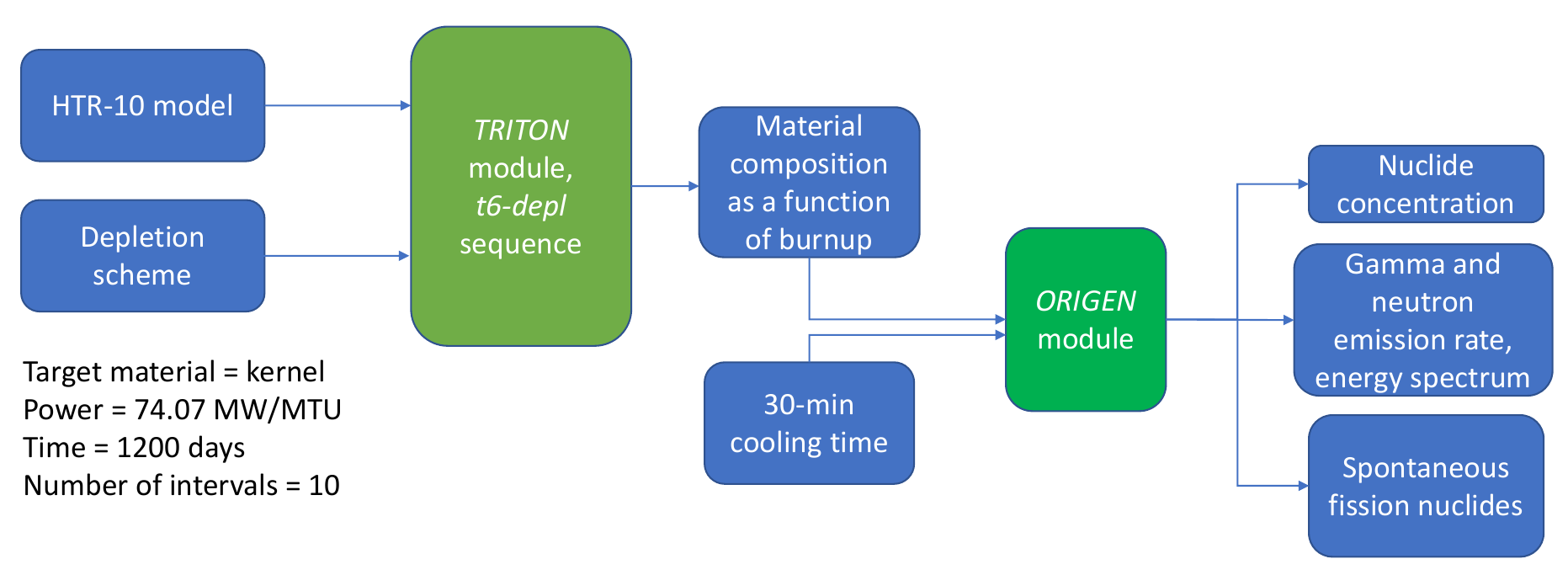}
	\caption{Software workflow to determine the pebble composition as a function of the burnup and the resulting source term for given cooling time periods.}
	\label{fig:scale_workflow}
\end{figure}

In this section, we used \software{SCALE} 6.2.4 to simulate the nuclide concentration of used fuel pebble.\added{ The
nuclear data library used was ENDF-VII.1.} SCALE is a widely-used modeling and simulation suite for nuclear safety analysis and design~\cite{rearden2018scale}. Figure~\ref{fig:scale_workflow} shows our simulation workflow. A high-fidelity HTR-10 reactor model was developed and a fuel depletion scheme were input to the \textit{TRITON} (Transport Rigor Implemented with Time-dependent Operation for Neutronic depletion) module of SCALE to calculate the fuel composition at a set of burnup levels. \textit{ORIGEN} (Oak Ridge Isotope Generation) module was then used to calculate the fuel isotopics, gamma-ray and neutron source terms, based on which simulations of neutron interrogation of used TRISO-fueled pebbles were performed in Section~\ref{sec:active_interrogation}. 
\subsection{SCALE Model of HTR-10}

\begin{figure}[!htbp]
    \captionsetup{font=footnotesize}
    \begin{subfigure}[t]{0.5\linewidth}
        \centering
    	\includegraphics[width=\linewidth]{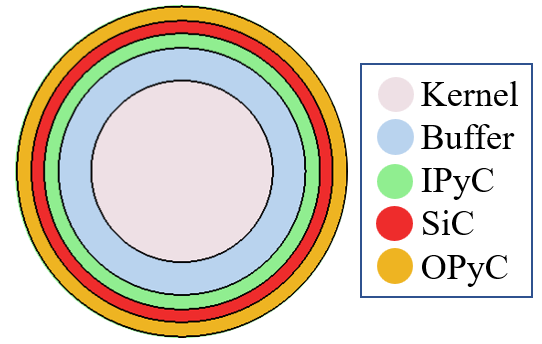}
    	\caption{Particle}
    	\label{fig:triso_prtl_scale}
    \end{subfigure}\hfil
    \begin{subfigure}[t]{0.5\linewidth}
        \centering
    	\includegraphics[width=.7\linewidth]{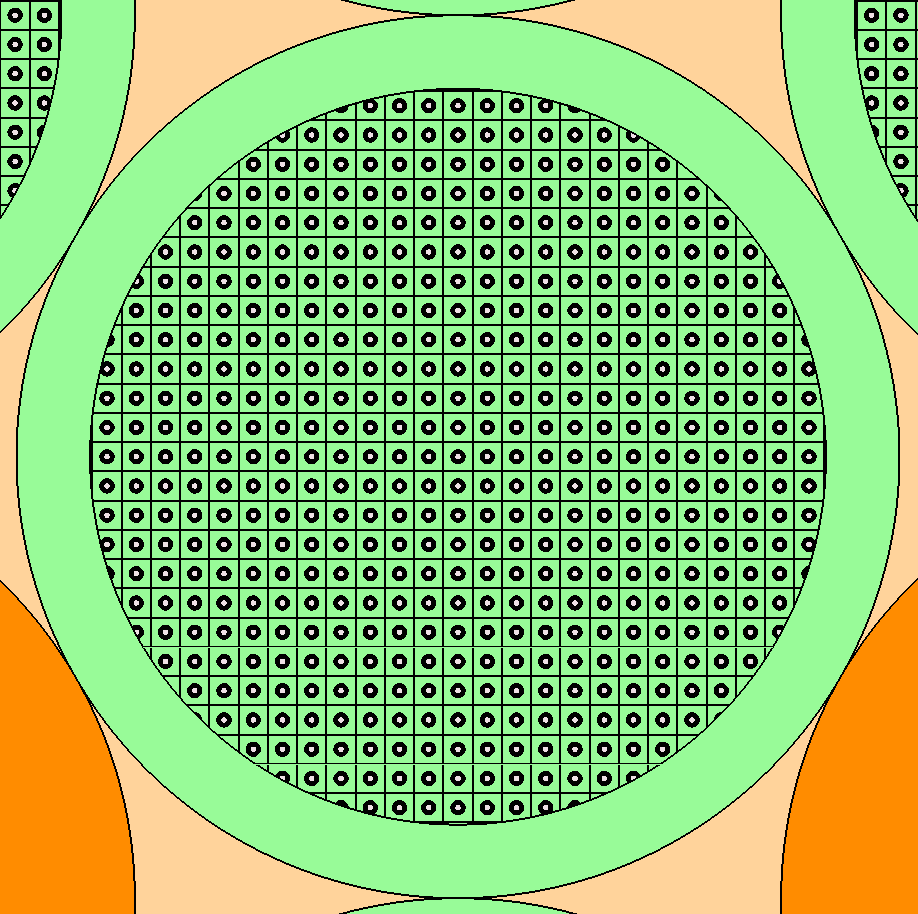}
    	\caption{Pebble}
    	\label{fig:triso_pebble_scale}
    \end{subfigure}
    \caption{SCALE model of a TRISO-fuel particle and a TRISO-fueled pebble. Black dots in b) represent TRISO fuel particles filled in the pebble.}
    % \label{fig:active_interrogation_results}
\end{figure}

As a first step of the depletion calculation, a Monte Carlo model of the {HTR-10} test reactor was developed in SCALE 6.2.4. HTR-10 is a 10-MW\textsubscript{th} PBR test reactor in China that started operation in 2000~\cite{xu2002htr}. Benchmark experiments were performed on HTR-10 under initial-critical and fully loaded conditions, which enables validation of our computational model~\cite{terry2005evaluation}.
Figure~\ref{fig:triso_prtl_scale} shows the cross section of a single TRISO particle in the HTR-10 model. A \SI{500}{\micro\meter} diameter UO\textsubscript{2} kernel is located at the center. The fuel kernel is then coated by a layer of \SI{90}{\micro\meter} thick low-density pyrolytic carbon (PyC) buffer layer to provide space for fission gas. It is then coated with three successive layers of inner PyC (IPyC, \SI{40}{\micro\meter} thick), SiC (\SI{35}{\micro\meter}) and outer PyC (OPyC, \SI{40}{\micro\meter}). The SiC is the main structural support and provides primary retention of non-gaseous fission products while the IPyC and OPyC provide additional fission gas retention~\cite{demkowicz2019triso,barrachin2010high}. The overall diameter of a TRISO fuel particle is \SI{910}{\micro\meter}. The \textsuperscript{235}U enrichment of the fuel is 17~wt\%~\cite{terry2005evaluation}. %Table~\ref{table:triso_particle_isotopic_composition} summarizes the material compositions of the kernel and coating layers.

Fuel elements in the HTR-10 reactor are spherical pebbles filled with TRISO fuel particles. Each pebble contains \SI{5}{\gram} uranium by design~\cite{terry2005evaluation}. The diameter of the pebble is 6~cm, and the diameter of the fuel zone is 5~cm. Previous studies show that the use of homogenized fuel zone model results in significant underestimation of the effective multiplication factor(k\textsubscript{eff})~\cite{colak2005monte} and hence individual TRISO particles need to be modeled explicitly. k\textsubscript{eff} is less sensitive to the distribution of the particles inside the pebble~\cite{colak2005monte}, therefore, we adopted a uniform distribution, instead of a random one. Figure~\ref{fig:triso_pebble_scale} shows the cross section of a TRISO-fueled pebble in the SCALE model. We created a body-centered cubic (BCC) lattice to populate the TRISO fuel particles in the fuel zone. The side length of the lattice cell is 0.19876~cm, and the number of layers in the X/Y/Z direction is 27~\cite{international2004iaea}. Taking the clipping by the boundary into account, the effective number of particles within the fuel zone is 8329.89, and the total mass of uranium in a pebble is \SI{4.997}{\gram}. The density of graphite matrix is \SI{1.73}{\g\per\cubic\cm} and the boron content concentration of the graphite matrix is 1.3~ppm~\cite{terry2005evaluation}. Apart from the TRISO-fueled pebbles, there are thousands of \SI{6}{\cm} diameter pure graphite pebbles in the core to moderate the neutrons. The density of graphite in pure graphite pebbles is \SI{1.84}{\gram\per\cubic\cm} and the boron content concentration is 0.125~ppm~\cite{terry2005evaluation}. %Table~\ref{table:triso_graphite_isotopic_composition} summarizes the material compositions of the graphite in TRISO-fueled pebbles and pure graphite pebbles.
% \begin{figure}[!htbp]
% 	% \captionsetup{font=footnotesize}
% 	\centering
% 	\includegraphics[width=.5\linewidth]{figs/scale_pebble.png}
% 	\caption{SCALE model of a TRISO-fueled pebble with black dots representing the TRISO fuel particles.}
% 	\label{fig:triso_pebble_scale}
% \end{figure}
% \begin{table}[!htbp]
% 	\centering
% 	\caption{Isotopic concentrations of graphite in TRISO-fueled pebbles and pure graphite pebbles (atoms/barn-cm).}\label{table:triso_graphite_isotopic_composition}
% 	\begin{tabular}{|c|c|c|c|c|}
% 		\hline
% 		       & Graphite in TRISO pebble         & Graphite in graphite pebble     \\ \hline
% 		\textsuperscript{10}B   & 2.49302659E-08 & 2.54956032E-09  \\ \hline
% 		\textsuperscript{11}B   & 1.00347452E-07 & 1.02623006E-08 \\ \hline
% 		Carbon & 8.67417390E-02 & 9.22572180E-02  \\ \hline
% 	\end{tabular}
% \end{table}

%Model of pebble region
\begin{figure}[!htbp]
	% \captionsetup{font=footnotesize}
	\centering
	\includegraphics[width=\linewidth]{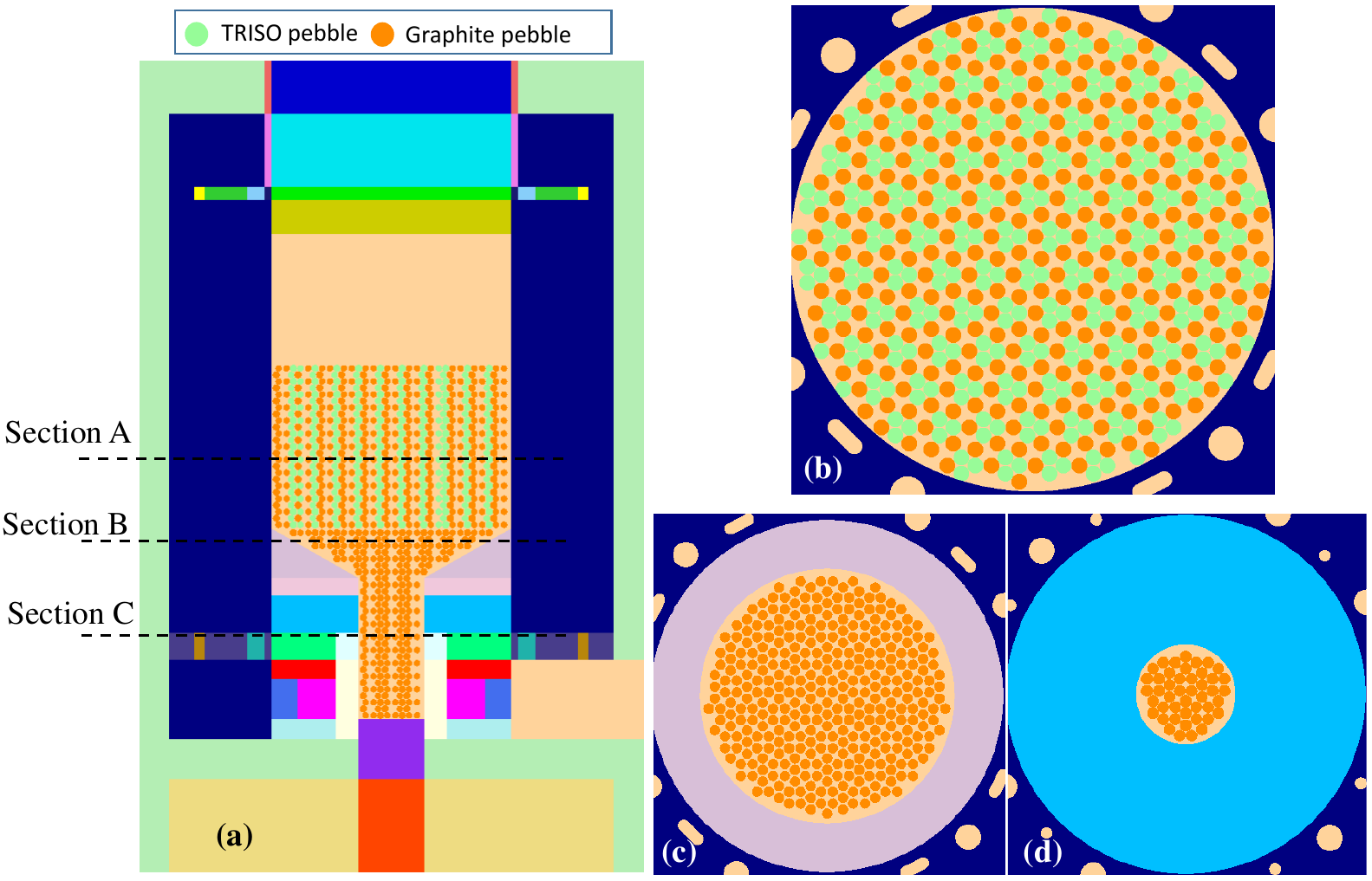}
	\caption{HCP lattice in the mixed pebble region, conus region and discharge tube. (a): Vertical cross section of the HTR-10 model. Green spheres represent the TRISO-fueled pebbles and the orange spheres represent the pure graphite pebbles. (b): Horizontal cross section of the mixed pebble region at Section A. (c): Horizontal cross section of the conus region at Section B. (d): Horizontal cross section of the discharge tube at Section C.}
	\label{fig:hcp_filling}
\end{figure}
\begin{figure}[!htbp]
	% \captionsetup{font=footnotesize}
	\centering
	\includegraphics[width=\linewidth]{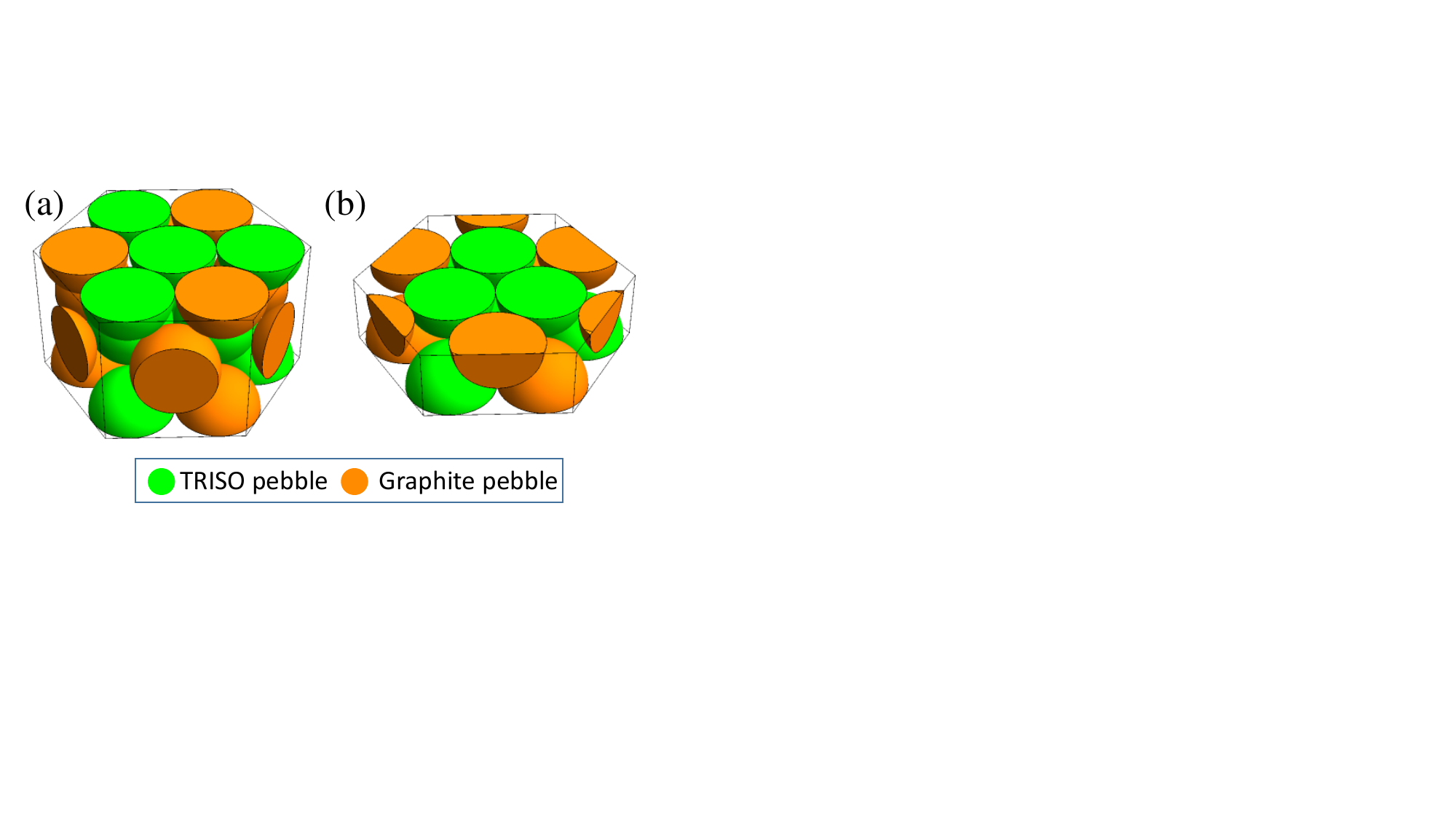}
	\caption{HCP lattice cell in the mixed pebble region. (a) A single HCP cell. (b) Bottom half of the HCP cell. Orange sphere represents the pure graphite pebble and green sphere represents the TRISO-fueled pebble.}
	\label{fig:hcp_lattice_cell}
\end{figure}
TRISO-fueled pebbles and pure graphite pebbles are present in three regions in the reactor, namely the mixed-pebble region (Fig.~\ref{fig:hcp_filling}b), the conus region(Fig.~\ref{fig:hcp_filling}c), and the discharge tube (Fig.~\ref{fig:hcp_filling}d). The mixed pebble-region is a cylinder of \SI{90}{\cm} radius and \SI{221.818}{\cm} height, which is filled with 9627 TRISO-fueled pebbles and 7263 pure graphite pebbles when the initial critical condition is reached~\cite{terry2005evaluation}. The equivalent fuel loading height of the core is \SI{123.06}{\cm} and the packing fraction is 61\%. The conus region is a truncated cone with \SI{90}{\cm} radius in the upper surface, \SI{25}{\cm} radius in the bottom surface, and \SI{36.946}{\cm} height. The discharge tube is a \SI{25}{\cm} radius and \SI{106.236}{\cm} tall cylinder. Both the conus region and the discharge tube are filled with pure graphite pebbles only. We used a Hexagonal Close-Packed (HCP) lattice to populate the pebbles in these regions~\cite{sunny2010scale,ilas2010scale}. Figure~\ref{fig:hcp_lattice_cell} shows a single HCP lattice cell in the mixed pebble region, which is a hexagonal prism of \SI[parse-numbers = false]{6+2\sqrt{3}}{\cm} side length and \SI[parse-numbers = false]{4\sqrt{6}}{\cm} height, with three layers of pebbles. The top and bottom layers contain three graphite hemispheres and four TRISO hemispheres each. The middle layer consists of three TRISO-fueled pebbles at the center and six graphite pebbles clipped by the cell's side surfaces. All pebbles are of \SI{6}{\cm} diameter and are tangent to each other. The HCP lattice cell in the conus region and discharge tube is of the same dimension but with all TRISO-fueled pebbles replaced by graphite pebbles. \replaced{We}{To make the model more realistic, we} have developed a custom python program to fill the core with the HCP cells\replaced{. Any}{ and remove any} pebbles that intersect with or lie outside of the core's boundary\added{ were removed because they would result in unrealistic and significant increase in k\textsubscript{eff}.} Figure~\ref{fig:hcp_filling} shows the cross sections of the optimized HCP lattice in the mixed pebble region, the conus region and discharge tube. % There are 14 layers of pebbles in the mixed pebble region and 16 layers of pebbles in the conus region\&discharge tube. 
By carefully tuning the arrangement of pebbles, we achieved the same number of TRISO-fueled pebbles in the mixed-pebble region compared to the experiment, seven extra graphite pebbles and 0.23\% lower packing fraction, as shown in Table~\ref{table:pebble_filling_comp_scale}, which are similar to the values reported in~\cite{sunny2010scale}.

\begin{table}[!htbp]
	\centering
	\caption{Comparison of pebble filling in the mixed pebble region between the SCALE model and experiment~\cite{terry2005evaluation}.}\label{table:pebble_filling_comp_scale}
	\begin{tabular}{|c|c|c|c|}
		\hline
		                 & Experiment & SCALE model & Difference \\ \hline
		TRISO pebbles    & 9627       & 9627        & 0          \\  \hline
		Graphite pebbles & 7263       & 7270        & +7         \\  \hline
		Core height      & 123.06 cm  & 123.57 cm   & +0.51 cm   \\  \hline
		Packing fraction & 61\%       & 60.77\%     & -0.23\%    \\  \hline
	\end{tabular}
\end{table}

%Model of the whole reactor
The pebbles are surrounded with graphite reflectors and borated carbon bricks to shield the thermal neutrons, as shown in \added{Fig.~\ref{fig:hcp_filling}a}. The material compositions of the surrounding graphite cells are listed in~\cite{terry2005evaluation}. Additional structures including coolant flow channels, control rod and irradiation channels, absorber channels and gas duct were also modeled based on the specifications in~\cite{terry2005evaluation}. \added{Figure~\ref{fig:hcp_filling}a} shows the full HTR-10 model in SCALE. The overall radius and height of HTR-10 are 190~cm and 610~cm, respectively. We used the \textit{KENO-VI} module in SCALE to calculate the the effective multiplication factor. The nuclear data library used was ENDF-VII.1 and the number of generations was 500 with 1000 neutrons per generation. An effective multiplication factor of $1.0008\pm0.0013$ was obtained, which is in good agreement with the experimental value of $1.00000\pm0.00369$~\cite{terry2005evaluation}.\added{ This result was obtained thanks to the detailed documentation on the dimensions and material compositions of the components in HTR-10~\cite{terry2005evaluation}, which was accurately reproduced in our SCALE model. Importantly, we accurately adjusted the pebble arrangement in the core to match the experimental number of pebbles and void faction as much as possible.} The excellent agreement of k\textsubscript{eff} between the simulation and experiment suggests that our Monte Carlo model of the HTR-10 is accurate, based on which we can perform depletion calculation of the fuel and extract the isotopic composition of partially spent fuel.
% \begin{figure}[!htbp]
% 	% \captionsetup{font=footnotesize}
% 	\centering
% 	\includegraphics[width=\linewidth]{figs/full_htr_scale.pdf}
% 	\caption{Full SCALE model of HTR-10. (a): Vertical cross-section of the reactor. Green spheres represent the TRISO-fueled pebble and the orange spheres represent the pure graphite pebbles. (b): Zoomed-in view of the mixed pebble region. (c): Zoomed-in view of a single TRISO-fueled pebble, with each black dot representing one TRISO fuel particle.}
% 	\label{fig:full_htr_scale}
% \end{figure}

\subsection{Fuel Burnup and Depletion Calculation}
Based on the validated HTR-10 model, the \textit{t6-depl} sequence in {TRITON} module of SCALE was employed to calculate the material composition at different burnup levels. When equilibrium fuel cycle is reached, the HTR-10 core contains 27,000 fuel pebbles~\cite{xu2002htr}, i.e., \SI{0.135}{\tonne} uranium, and the thermal power is \SI{10}{\mega\watt}. Therefore, the reactor power in simulation was set to 74.07~MW/MTU. The depletion time was 1200 days in ten time intervals and the resulting fuel burnup ranged from \SI{8.88}{\gwdpert} to \SI{88.88}{\gwdpert} in steps of \SI{8.88}{\gwdpert}. The designed fuel burnup value of HTR-10 is \SI{80}{\gwdpert}~\cite{yang2002fuel} and our depletion scheme covers the full range. The fuel material composition at each burnup step was then fed to the {ORIGEN} module of SCALE to compute the nuclide concentration, gamma-ray source term, and neutron source terms, including the \replaced{($\alpha$,n)}{($\alpha$, n)} neutron, delayed neutron, and spontaneous fission neutron.
\begin{figure}[!htbp]
	\centering
	\includegraphics[width=\linewidth]{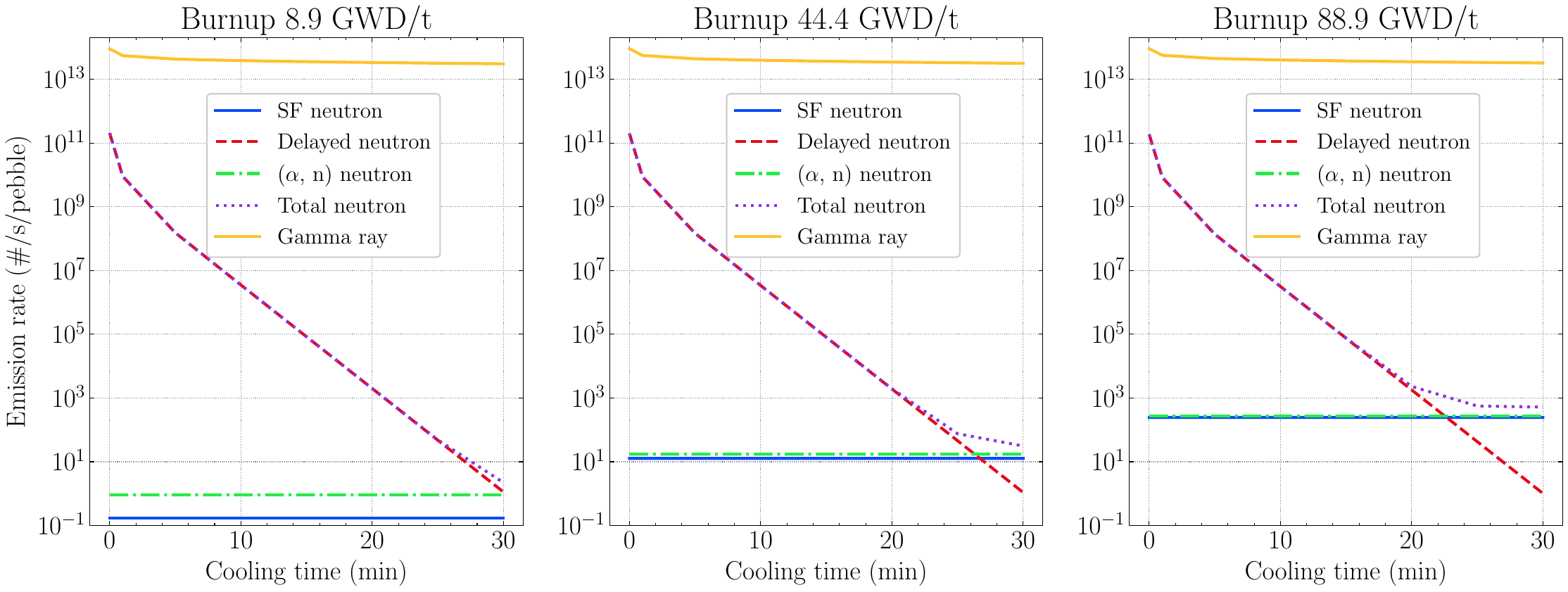}
	\caption{\replaced{Simulated gamma-ray}{Gamma-ray}, \replaced{($\alpha$,n)}{($\alpha$, n)} neutron, delayed neutron, SF neutron and total neutron emission rate as a function of cooling time at burnup = 8.9, 44.4, 88.9~GWd/t}.
	\label{fig:emission_rate}
\end{figure}

Figure~\ref{fig:emission_rate} shows the gamma-ray and neutron emission rate as a function of fuel cooling time at different burnup levels. Initially, the delayed neutron is of high intensity and may saturate the detector in neutron measurements. After a cooling period of approximately \SI{30}{\min}, the delayed neutron count rate decreases to approximately the same level as \replaced{($\alpha$,n)}{($\alpha$, n)} and spontaneous fission neutron rates due to the decay of delayed neutron precursors. Although the cooling time varies depending on the specific reactor resign~\cite{SU2006686}, tens of hours of cooling is typically required before performing gamma-ray spectroscopy measurements to determine the burnup~\cite{CHEN2003393,YAN2014172,VERGARI2021111189}. Therefore, a cooling time of \SI{30}{\min} is reasonably short and compatible with reactor operation. The nuclide concentration after \SI{30}{\min} of decay was used in the following calculation. Figure~\ref{fig:emission_rate} also shows that the gamma-ray intensity is approximately 12 orders of magnitude higher than neutrons at the end of cooling. A low gamma-ray sensitivity of 10\textsuperscript{-12} of the neutron counting system is hence required to reject counts from gamma rays.
\begin{figure}[!htbp]
	\centering
	\includegraphics[width=.7\linewidth]{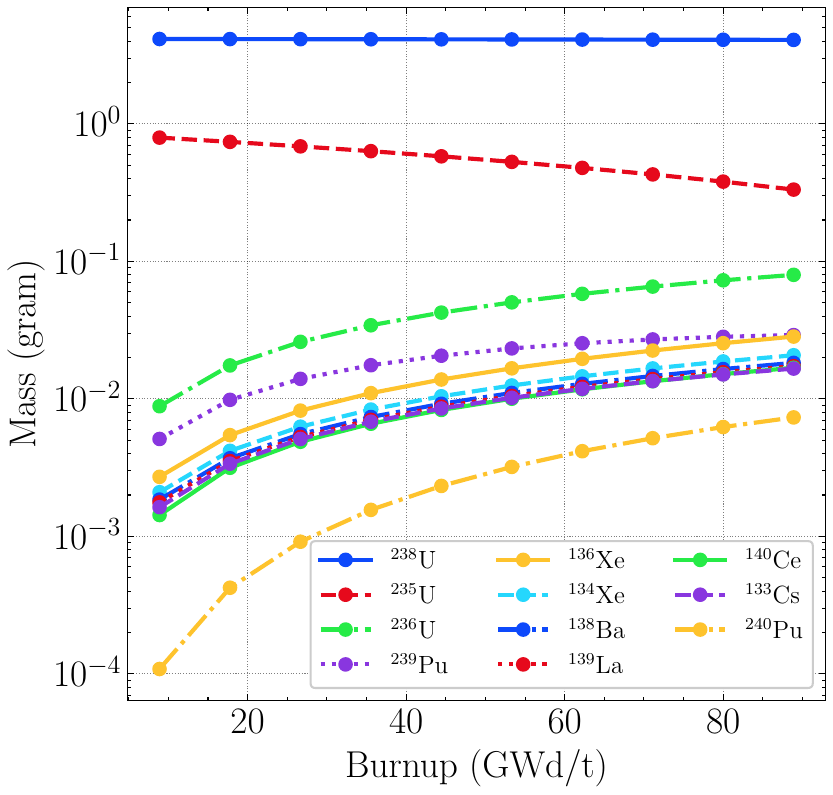}
	\caption{\replaced{Simulated mass}{Mass} of 10 nuclides of the highest weight fraction.}
	\label{fig:nuclide_mass}
\end{figure}
\begin{figure}[!htbp]
	\centering
	\includegraphics[width=.7\linewidth]{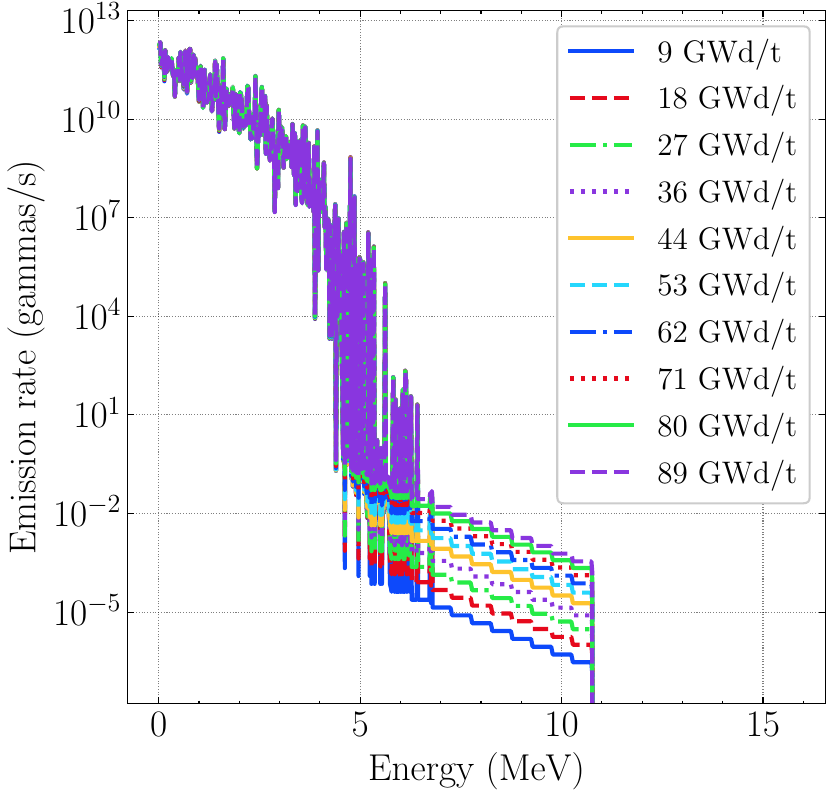}
	\caption{\replaced{Simulated gamma-ray}{Gamma-ray} emission spectrum as function of burnup.}
	\label{fig:gamma_spectra}
\end{figure}

\begin{figure}[!htbp]
	\centering
	\includegraphics[width=.7\linewidth]{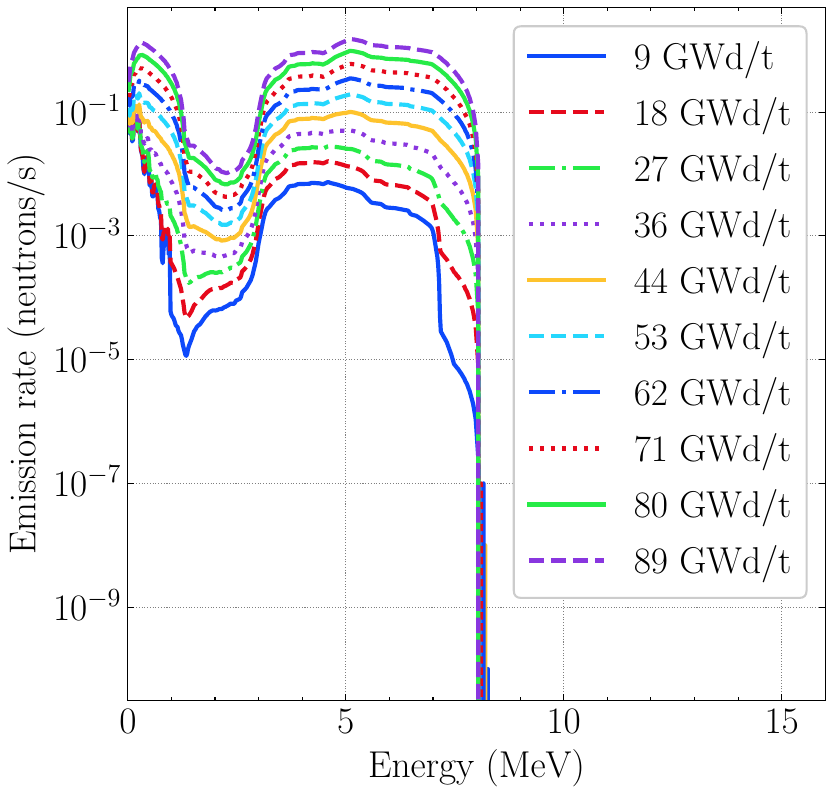}
	\caption{\replaced{Simulated energy}{Energy} spectrum of delayed and \replaced{($\alpha$,n)}{($\alpha$, n)} neutrons as function of burnup.}
	\label{fig:delayed_alpha_spectra}
\end{figure}
\begin{figure}[!htbp]
	\centering
	\includegraphics[width=.7\linewidth]{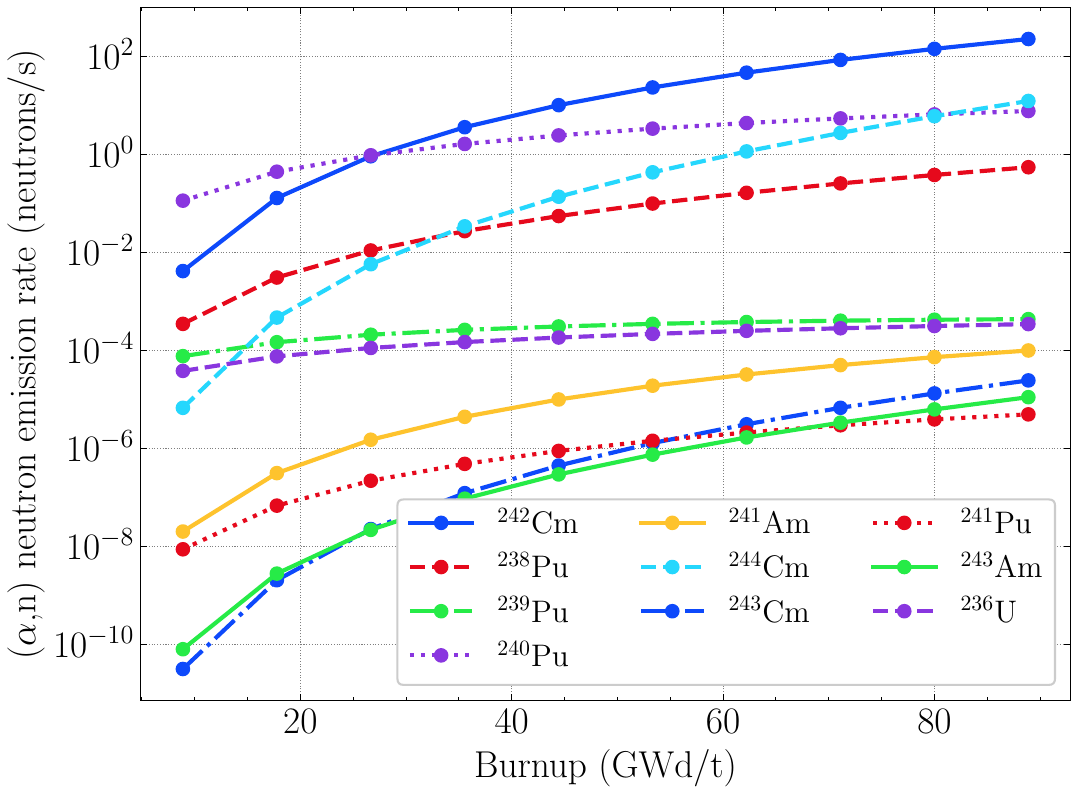}
	\caption{\added{Simulated} ($\alpha$,n) neutron rate of the top-10 contributors as function of burnup.}
	\label{fig:alpha_n_neutron_intensity}
\end{figure}
\begin{figure}[!htbp]
	\centering
	\includegraphics[width=.7\linewidth]{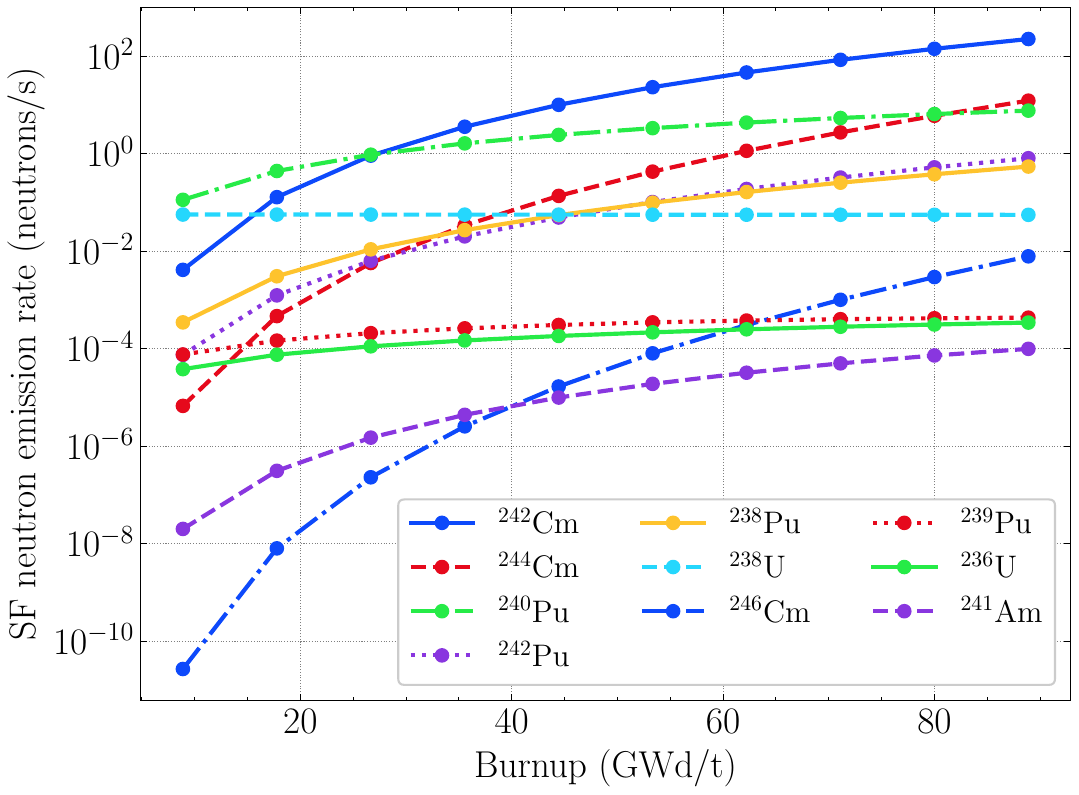}
	\caption{\added{Simulated} SF neutron rate of the top-10 SF neutron emitters as function of burnup.}
	\label{fig:sf_neutron_intensity}
\end{figure}

Figure~\ref{fig:nuclide_mass} shows the mass of ten nuclides of the highest weight fraction as a function of burnup calculated by ORIGEN. The \textsuperscript{235}U mass decreases and the mass of fission products increases with burnup. Figure~\ref{fig:gamma_spectra} shows the gamma-ray emission spectrum as a function of burnup. Figure~\ref{fig:delayed_alpha_spectra} shows the delayed and \replaced{($\alpha$,n)}{($\alpha$, n)} neutron emission spectra as a function of burnp, both acting as un-correlated neutron source term in the simulation in Section~\ref{sec:active_interrogation}. A few delayed neutron precursors such as \textsuperscript{98}Rb and \textsuperscript{136}Sb can create multiple delayed neutrons at the same time~\cite{LIANG20201}. However, most delayed neutron precursors have half-lives below \SI{1}{\s} and after the \SI{30}{\min} cooling period, the only delayed neutron precursor of considerable amount is \textsuperscript{87}Br, which has a half-life of \SI{55.64}{\s}~\cite{LIANG20201}. Only one delayed neutron is emitted after the beta-decay of \textsuperscript{87}Br, in the following reaction:
\begin{linenomath*}
\begin{gather*}
    {}^{87}\mathrm{Br} \to {}^{87}\mathrm{Kr} + e^{-} + \Bar{\nu}\\
    {}^{87}\mathrm{Kr} \to {}^{86}\mathrm{Kr} + n
\end{gather*}
\end{linenomath*}
Therefore, the delayed neutrons can be treated as uncorrelated in time.\added{ Fig.~\ref{fig:alpha_n_neutron_intensity} shows the ($\alpha$,n) neutron emission rate by nuclide.} The \replaced{($\alpha$,n)}{($\alpha$, n)} neutrons are mostly due to the following two reactions~\cite{SHURSHIKOV1985509,cristallo2018importance} and are also uncorrelated in time:
\begin{linenomath*}
\begin{gather*}
    {}^{242}\mathrm{Cm} \to \begin{cases*}
                    {}^{238}\mathrm{Pu} + \alpha (6.069 \mathrm{MeV}) & 25.92\%    \\
                    {}^{238}\mathrm{Pu} + \alpha (6.113 \mathrm{MeV}) & 74.08\% 
                 \end{cases*}\\
    {}^{13}\mathrm{C} + \alpha \to {}^{16}\mathrm{O} + n + 2.22 \mathrm{MeV}
\end{gather*}
\end{linenomath*}
The maximum energy of \replaced{($\alpha$,n)}{($\alpha$, n)} neutrons released through these two reactions is approximately \SI{8}{\mega\eV}, shown as the cutoff in Fig.~\ref{fig:delayed_alpha_spectra}.\added{ It should be pointed out that we assumed the pebble material to be homogeneous in the ORIGEN calculation, which made \textsuperscript{13}C($\alpha$,n)\textsuperscript{16}O reaction almost the only ($\alpha$,n) neutron contributor. However, in reality, some alphas reacted with \textsuperscript{17}O and \textsuperscript{18}O in the kernel, which had smaller macroscopic reaction cross-sections compared to \textsuperscript{13}C in the homogenized material (\SI{5.12e-5}{\per\cm} for \textsuperscript{17}O($\alpha$, n)\textsuperscript{20}Ne plus \textsuperscript{18}O($\alpha$, n)\textsuperscript{21}Ne, \SI{2.33e-4}{\per\cm} for \textsuperscript{13}C($\alpha$, n)\textsuperscript{16}O) and resulted in fewer ($\alpha$,n) neutrons. Therefore, the ($\alpha$,n) neutron emission rate calculated here and the resulting count rate from ($\alpha$,n) neutrons in Section~4 were conservatively overestimated, though still negligible.} The \replaced{($\alpha$,n)}{($\alpha$, n)} neutron intensity increases with the burnup as the amount of \textsuperscript{242}Cm increases.\added{ The sum of the delayed and ($\alpha$,n) neutrons can therefore be treated as a single effective uncorrelated neutron source term, hereafter referred to as ``the delayed \& ($\alpha$,n) neutron source''.} Figure~\ref{fig:sf_neutron_intensity} shows SF neutron intensity contributed by the top-10 SF neutron emitters, all increasing with the fuel burnup. The major SF neutron contributors are \textsuperscript{242}Cm, \textsuperscript{244}Cm, and \textsuperscript{240}Pu, which acted as the correlated neutron source term in the simulation in Section~\ref{sec:active_interrogation}.

\section{Experimental Method}\label{sec:NMC_description}
\subsection{Introduction}
The neutronics and depletion fuel analysis of a prototypical PBR allowed us to determine the constraint that the NMC needs to meet to be able to measure \textsuperscript{235}U mass in used TRISO pebbles. In this section, we describe the instrumentation and experimental approach to measure the \textsuperscript{235}U in TRISO-fueled pebbles through neutron coincidence counting and demonstrate the gamma-ray insensitivity of the BCS-based NMC.
\subsection{BCS-based Neutron Multiplicity Counter}
\begin{figure}[!htbp]
	\centering
	\includegraphics[width=\linewidth]{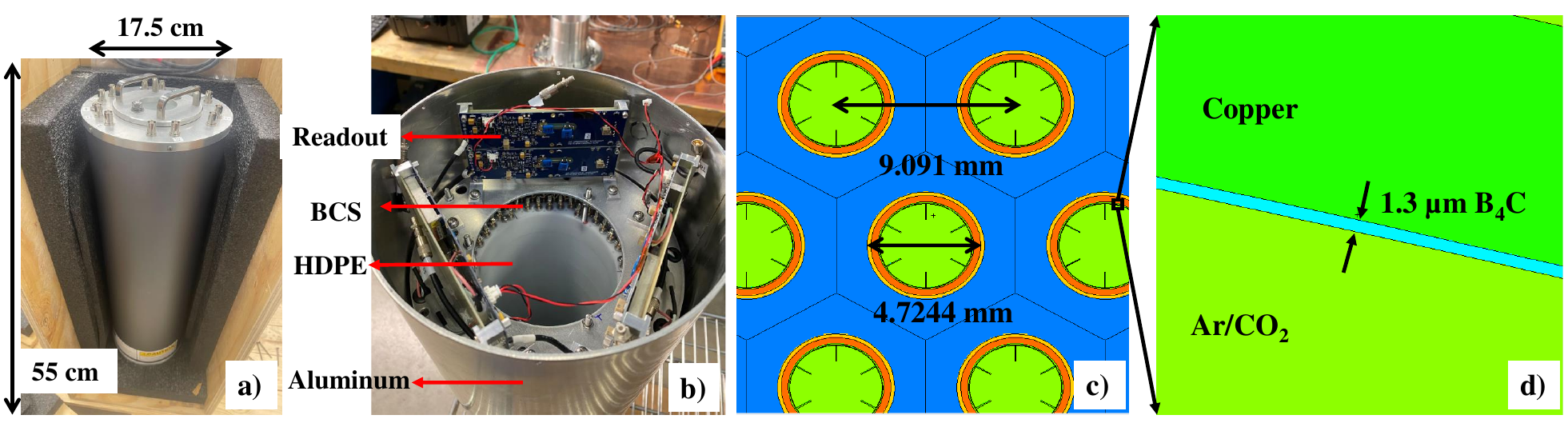}
	\caption{Design of a custom BCS-based NMC for interrogation of TRISO-fueled pebbles. (a) is the external view of the counter, (b) is the internal view with the readout electronics, straws and HDPE, (c) is the cross section of pie-6 straws, { and (d) shows} the \SI{1.3}{\micro\meter}-thick layer of B\textsubscript{4}C converter.}
	\label{fig:NMC_mcnp_model}
\end{figure}

 Figure~\ref{fig:NMC_mcnp_model} shows our design of \added{a prototype }of a custom BCS-based NMC for interrogation of TRISO-fueled pebbles. The NMC is a cylinder that measures \SI{55}{\cm} in length and \SI{17.5}{\cm} in diameter, as shown in Fig.~\ref{fig:NMC_mcnp_model}a. It has a central cavity with a diameter of \SI{10}{\cm}, which serves as the sample placement area, as shown in Fig.~\ref{fig:NMC_mcnp_model}b. The inner and outer surfaces of the NMC are covered with 0.508 mm"~thick cadmium to prevent thermal neutrons from reentering the sample cavity. The NMC \added{prototype }is composed of 192 straws that are organized in a hexagonal lattice, with an inter-straw distance of \SI{0.9091}{\cm}, as shown in Fig.~\ref{fig:NMC_mcnp_model}c. Surrounding the straws is high-density polyethylene. Each straw is \SI{40}{\cm} long and has a diameter of \SI{4.7244}{\mm}. Figure~\ref{fig:NMC_mcnp_model}d shows that the inner surface of each straw is coated with a \SI{1.3}{\micro\meter}-thick layer of B\textsubscript{4}C (96\% enriched with \textsuperscript{10}B) to absorb thermal neutrons and generate \textsuperscript{7}Li ions and alpha particles through ${}^{10}$B(\textsuperscript{1}n,$^4\alpha$)${}^{7}$Li reactions. Six septa are present inside each straw to increase the surface area for improving the detection efficiency and reducing the system die-away time. The straw is filled with a gas mixture of Ar/CO\textsubscript{2} (9:1) at 0.7 atmosphere to detect \textsuperscript{7}Li/alpha ions. When alpha/\textsuperscript{7}Li particles ionize the gas, ionized electron-ion pairs induce electrical signals on the anode wire at the center of the straw.\added{ Straws are read out in 6 groups of 32, each with its own custom-designed amplifier and neutron discriminator built into the NMC housing. A TTL pulse is produced if the signal is above the discriminator threshold. Apart from the six TTL outputs, an additional analog output from one of the six detector groups is available for analog acquisition.}\added{ Experimental characterization and Monte Carlo modeling of the NMC were performed in our previous work~\cite{FANG2023109794}. The neutron efficiency and die-away time, measured with a \textsuperscript{252}Cf source, was 4.71\% and \SI{16.90}{\micro\second}, respectively. The dead-time of one channel measured with an Agilent DSO-X 2002A Oscilloscope was approximately \SI{180}{\ns}.}\added{ An extended version of the BCS-based NMC is currently under development, which consists of 396 straws and has an increased neutron efficiency of 13.23\% and die-away time of \SI{21.56}{\micro\second}.} 
 
 \added{The timestamps collected from all TTL channels were merged and sorted. } Based on the timestamps of acquired pulses, we can calculate the neutron singles, doubles and triples count rates using the signal-triggered shift-register algorithm according to the multiplicity counting theory~\cite{langner1998application,Croft2012152,santi2014china}. \deleted{The pre-delay time was set to 2~$\mu$s, the gate was set to 32~$\mu$s, and the long delay time was set to 2~ms.}\added{ Let $P_n$ and $Q_n$ be the number of $n$-multiples in the first gate after pre-delay and the second gate after long delay, respectively. The singles rate ($S$), doubles rate ($D$), and triples rate ($T$) are:} 
 \begin{equation}\label{eq:S_D_T}
     \begin{aligned}
         S &= \frac{1}{L}\sum_{n=0}^{\infty} Q_n\\
         D &= \frac{1}{L}\sum_{n=1}^{\infty} n(P_n-Q_n)\\
         T &= \frac{1}{L} \left[\sum_{n=2}^{\infty} \frac{n(n-1)}{2}(P_n-Q_n)-\frac{D}{S}\sum_{n=1}^{\infty} nQ_n\right]
     \end{aligned}
 \end{equation}
 \added{where $L$ is the measurement time. }The uncertainties associated with \replaced{$S, D, T$}{singles and doubles} are as derived by Prasad et al.~\cite{prasad2018analytical}:
\begin{equation}\label{eq:S_D_uncertainty}
\begin{array}{l}
\sigma_S=\sqrt{\frac{S+2 D x_2}{L}}\\
\sigma_D=\sqrt{\frac{D+2 S^2 G+2 T+4 D S G x_2+2 \frac{D^3 x_2}{S^2}}{L}}\\
\sigma_T=\sqrt{\begin{array}{l}
\frac{T+D^2 G+D S G+2 S T G+S^3 G^2+2 \frac{D T^2 x_2}{S^2}+10 D^2 G x_2}{L} \\
+\frac{2 \frac{D^3 G x_2}{S}+2 D S G x_2+4 D T G x_2+4 D S^2 G^2 x_2+4 D^2 S G^2 x_2^2}{L} \\
+\frac{12 D T G x_3+12 S T G x_3}{L}
\end{array}}
\end{array}
\end{equation}
\begin{equation}
    x_k = \frac{\sum_{j=0}^{k-1}\binom{k-1}{j}(-1)^j\frac{1-e^{-j \lambda G}}{j \lambda G}}{\left[e^{-\lambda T_{pd}}(1-e^{-\lambda G})\right]^{k-1}}
\end{equation}
\replaced{where $\lambda$ is the inverse of die-away time $\tau$, $T_{pd}$, $G$, $T_{ld}$ are the pre-delay, gate width, and long delay, respectively. Although the NMC can detect triples, in this work we limited our analysis up to the second order, i.e., $S$ and $D$, due to low statistics in $T$.}{where $S$ and $D$ are the singles and doubles count rates, respectively, $G$ is the time gate width, $f_1$ is the gate utilization factor, and $L$ is the interrogation time.}

\subsection{Demonstration of High-Gamma-Ray-Insensitivity of BCS-Based NMC}\label{sec:gamma_insensitivity}
\begin{figure}[!htbp]
	\centering
	\includegraphics[width=.5\linewidth]{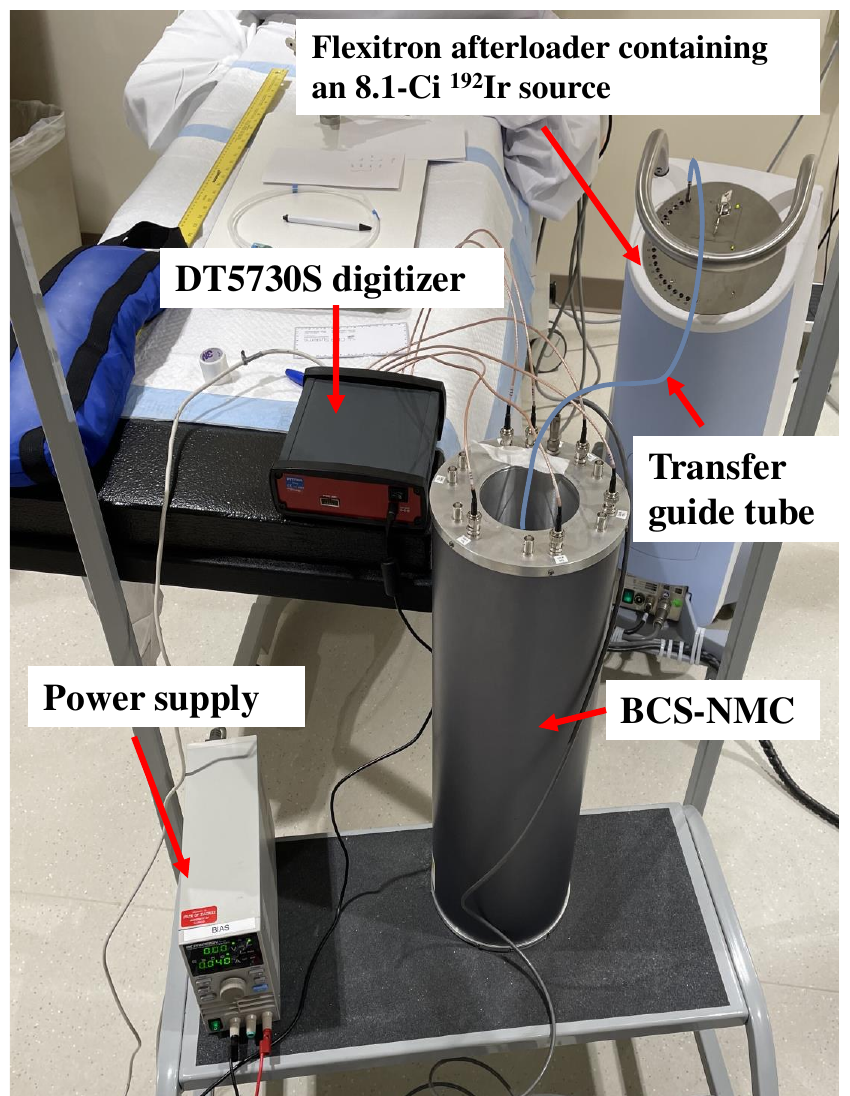}
	\caption{Measurement of the \SI{8.1}{\Ci}\added{ (\SI{2.997e11}{\becquerel})} \textsuperscript{192}Ir HDR source. The gamma-ray source was placed at the center of sample cavity of the NMC and the measurement lasted for \SI{10}{\min}.\added{ The blue transfer guide tube for transferring the source has a diameter of a few millimeters and is enlarged in the picture for visibility.}}
	\label{fig:ir-192-setup}
\end{figure}

\begin{figure}[!htbp]
	\centering
	\includegraphics[width=.7\linewidth]{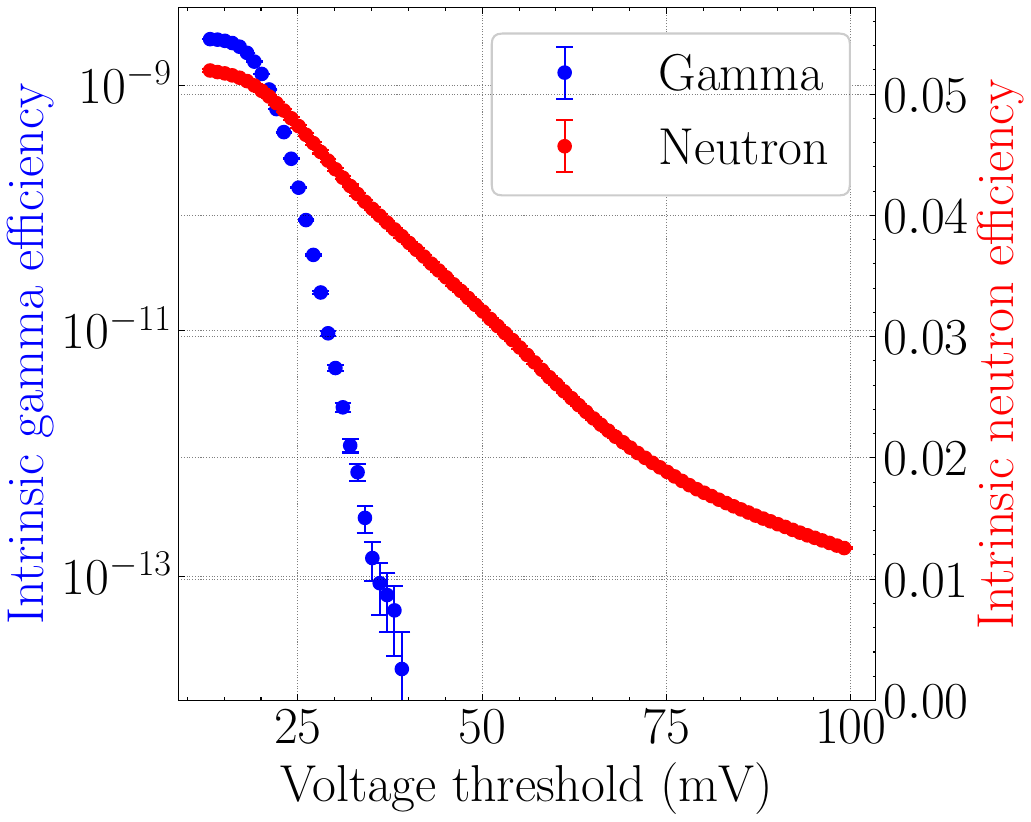}
	\caption{\replaced{Experimental intrinsic}{Intrinsic} gamma-ray and neutron efficiency \added{of the analog channel for singles }as a function of discriminator threshold.}
	\label{fig:gamma-efficiency}
\end{figure}

Used TRISO-fueled pebbles can emit passive gamma-rays with an intensity up to 10\textsuperscript{13} \replaced{gammas/s/pebble}{cps/pebble}~\cite{su2004design}. Therefore, gamma-ray-insensitivity of the NMC is crucial to ensure that passive gamma-rays emitted by the pebble do not interfere with neutron measurements. Gamma-rays produce low-amplitude pulses by depositing a small amount of energy, which can be effectively rejected by raising the discriminator threshold of the system~\cite{knoll2010radiation}. To demonstrate the system's high-gamma-ray insensitivity, we used our BCS-based NMC to measure a Nucletron Flexisource \textsuperscript{192}Ir HDR (High Dose-Rate) source in an Elekta Flexitron Remote Afterloader~\cite{afterloader}, as shown in Fig.~\ref{fig:ir-192-setup}.\added{ The gamma-ray fields of the \textsuperscript{192}Ir source and a spent fuel pebble with \SI{1}{\cm} tungsten shielding are comparable in terms of intensity, $6.4\times 10^{11}$~$\gamma$/s and $4\times 10^{12}$~$\gamma$/s, respectively. While the energy emitted by \textsuperscript{192}{Ir} is lower, the energy deposited by electrons within the straw would not be significantly different and the spectra were similar.} The \textsuperscript{192}Ir source has a diameter of \SI{0.6}{\mm} and length of \SI{3.5}{\mm}~\cite{lopes2022vivo}, with an activity of \SI{8.1}{\Ci}\added{ (\SI{2.997e11}{\becquerel})} at the date of experiment. The source can be transferred from the afterloader to the cavity of the NMC through a \replaced{transfer guide tube}{wire} and the source position can be controlled through a remote software. We placed the source at the center of sample cavity of the NMC and measured the source for \SI{10}{\min}. The number of gamma-rays impinging on the NMC's inner surface is $6.175\times10^{11}$ per second and the average exposure rate is \SI[per-mode = symbol]{340.87}{\Roentgen\per\hour}\added{ (\SI[per-mode = symbol]{8.3e-4}{\gray\per\s})}. The NMC was connected to a 14-bit 500MS/s DT5730S CAEN digitizer for list-mode data acquisition. The NMC was powered by a BK Precision 9110 power supply at \SI{5}{\V}. We calculated the intrinsic gamma-ray efficiency, defined as the ratio between the number of counts and the gamma-rays that reach the NMC's inner surface. Figure~\ref{fig:gamma-efficiency} shows the intrinsic gamma-ray efficiency as well as the intrinsic neutron efficiency obtained by measuring a \textsuperscript{252}{Cf} source~\cite{FANG2023109794} for various low-level threshold values. \added{The neutron pulse height distribution has a plateau starting from zero because of the energy loss of the alpha/\textsuperscript{7}Li ion in the \textsuperscript{10}B layer~\cite{FANG2023109794}. Therefore, the neutron efficiency strongly depends on the threshold value. }A gamma-ray efficiency of 10\textsuperscript{-12} is achievable\added{ at a threshold of \SI{32}{\mV}}, superior to typical \textsuperscript{3}He-based systems that feature 10\textsuperscript{-9} efficiency~\cite{KOUZES2011412,AKhaplanov_2013}, without sacrificing the neutron efficiency significantly.

\section{Neutron Interrogation of Partially Spent TRISO-Fueled 
Pebbles}\label{sec:active_interrogation}

\subsection{Introduction}
\begin{figure}[!htbp]
	\centering
	\includegraphics[width=\linewidth]{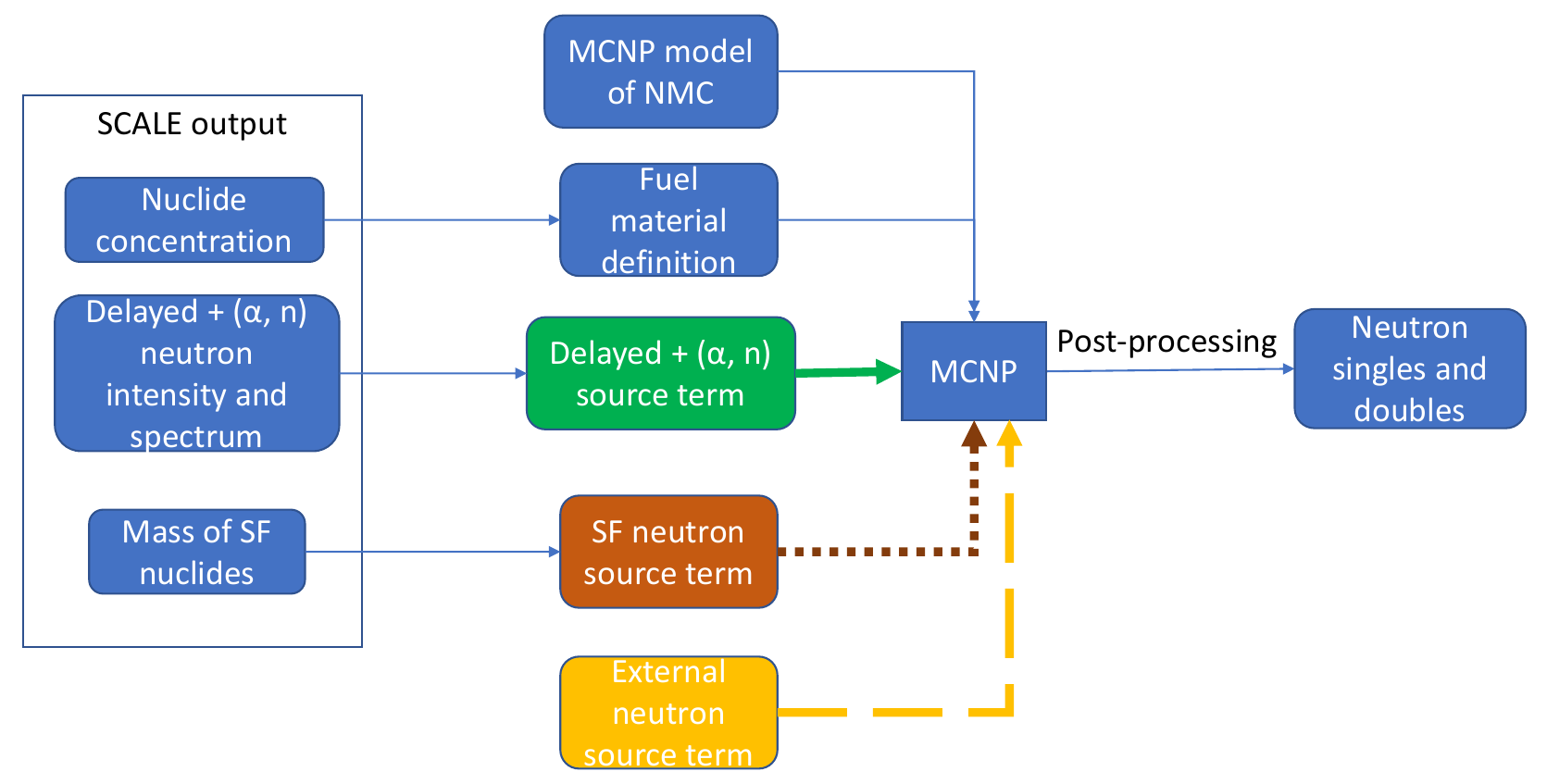}
	\caption{Software workflow to calculate the neutron singles and doubles count rates contributed by three neutron source terms separately.}
	\label{fig:mcnp_workflow}
\end{figure}

We simulated the neutron interrogation of a used fuel pebble at different burnup level using the BCS-based NMC in \textit{MCNP} (Monte Carlo N-Particle)~\cite{osti_1419730}. MCNP is a general-purpose Monte Carlo radiation transport code developed and maintained by Los Alamos National Laboratory. Figure~\ref{fig:mcnp_workflow} shows the simulation workflow. A high-fidelity model of the BCS-based NMC system was developed in MCNP and validated with a \textsuperscript{252}Cf measurement~\cite{FANG2023109794}. Our modeled singles and doubles count rate in passive mode agreed with experiment within 0.4\%. In active mode, the BCS-NMC exhibited a linear response of time-correlated neutron doubles when assaying samples of \textsuperscript{235}U mass in the 0.8-\SI{4.8}{\gram} range, comparable with \textsuperscript{235}U mass in TRISO-fueled pebbles. The experimentally validated model of the BCS-NMC was used to simulate the assay of fresh TRISO-fueled pebbles~\cite{FANG2023109794}. The TRISO fuel material composition at each burnup step was extracted from the SCALE simulation output. For each burnup, \replaced{we performed three MCNP simulations, with the source term being the ($\alpha$,n) neutron \& delayed neutron source, SF neutron source, and an external neutron interrogation source, respectively.}{we simulated the neutron singles and doubles count rates contributed by three neutron source terms, namely the {($\alpha$, n)} neutron and delayed neutron source, SF neutron source, and an external neutron interrogation source.} The list of timestamps of neutron pulses were extracted from the MCNP output file~\cite{FANG2023109794}, and a signal-triggered shift-register algorithm was used to compute the neutron singles and doubles count rates\added{ due to each source term to understand their relative contributions}~\cite{langner1998application,Croft2012152,santi2014china}.\added{ Finally, total singles and doubles count rates due to all source terms were calculated.} 

\subsection{\replaced{Comparison of singles and doubles rates from various source terms}{Active interrogation}}
\begin{figure}[!htbp]
	\centering
	\includegraphics[width=.7\linewidth]{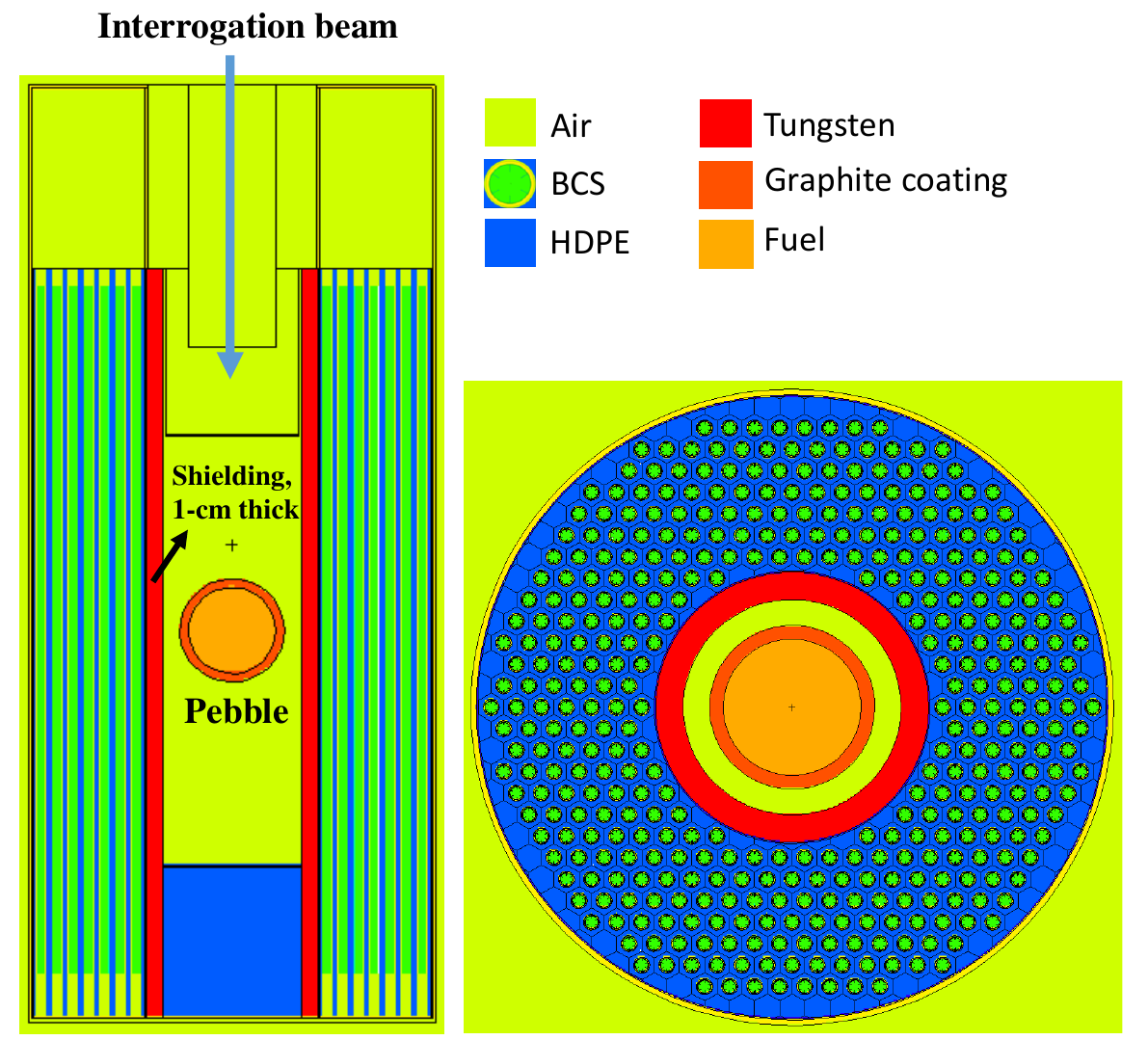}
	\caption{\replaced{MCNP simulation setup of active interrogation of a TRISO-fueled pebble with a thermal neutron beam. Left: Vertical cross section. Right: Horizontal cross section.}{Active interrogation measurement setup.}}
	\label{fig:active_interrogation_setup}
\end{figure}

\begin{figure}[!htbp]
	\centering
	\includegraphics[width=\linewidth]{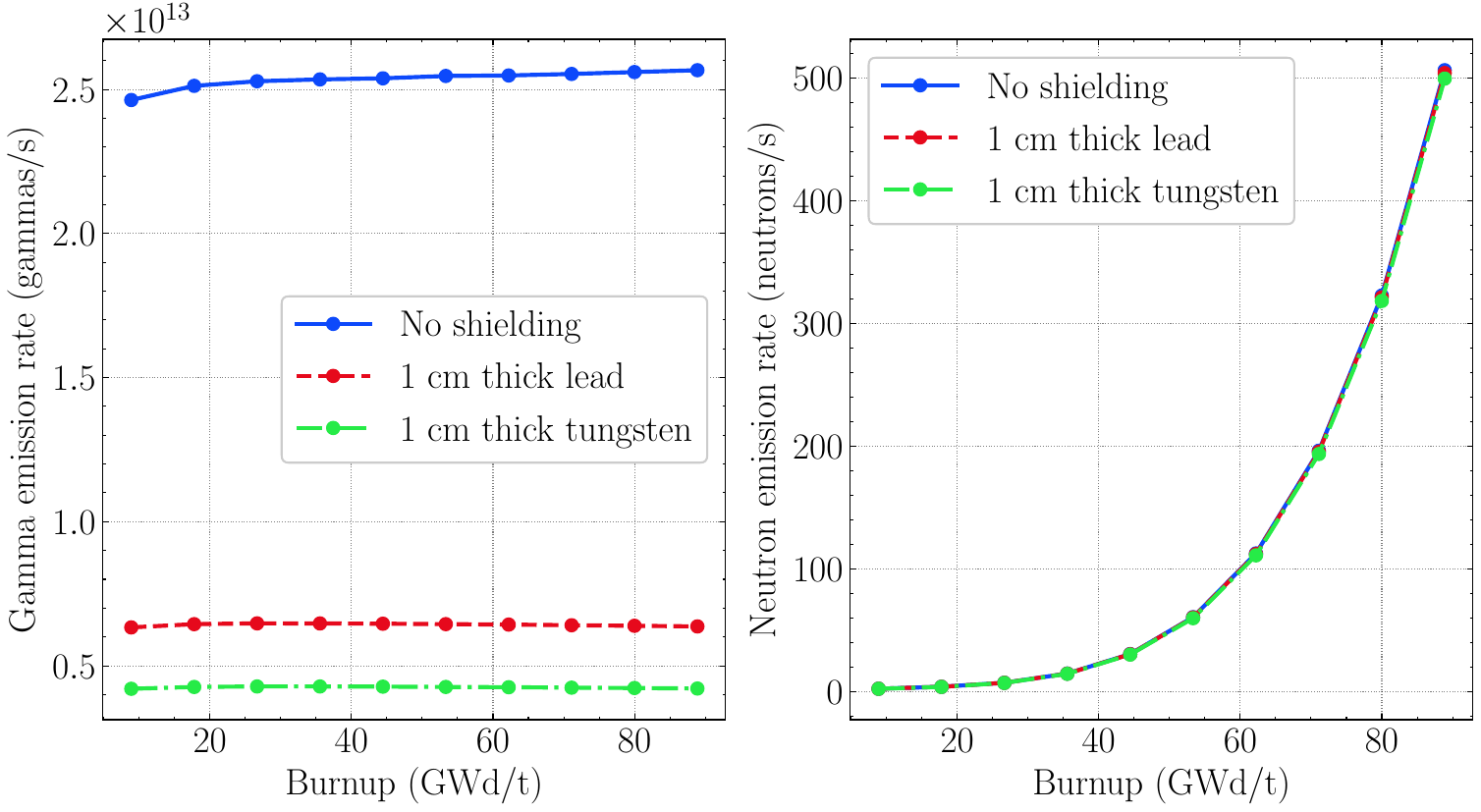}
	\caption{\replaced{Simulated gamma-ray}{Gamma-ray} and neutron emission rate after adding the shielding. \SI{1}{cm} thick tungsten shielded more gamma-rays than lead, and was used in the interrogation.}
	\label{fig:shielding}
\end{figure}

 \added{In active mode, we used an external neutron beam to induce fission in the sample.} Figure~\ref{fig:active_interrogation_setup} shows the MCNP simulation setup of interrogation of used TRISO-fueled pebbles with an external neutron source. A \SI{3}{\cm} radius TRISO-fueled pebble was placed at the center of the sample cavity of the NMC. A layer of \SI{1}{\cm} thick of tungsten was added to the inner surface of the NMC to shield the gamma-rays. MCNP simulation shows that \replaced{\SI{1}{\cm} of}{the} tungsten shielding reduces the gamma-ray intensity by approximately \replaced{84\%}{16\%} and neutron intensity by only 1.3\%, as shown in Fig.~\ref{fig:shielding}. The \replaced{measurement time}{interrogation time} was set to \SI{100}{s}. \replaced{In order to achieve a high assay accuracy within a short measurement time of \SI{100}{\s}, we need an active interrogation source that produces a high thermal neutron flux of {at least} $10^6$~n/cm$^2$/s and a low fraction of fast neutron background. In the simulation, the interrogation source was a collimated thermal neutron beam with an aperture of \SI{1}{\cm} diameter and a source strength of $10^6$~n/s, corresponding to a thermal neutron flux of $1.27\times10^6$~n/cm$^2$/s. These interrogation beam requirements can be met by several reactor beam ports and also by emerging commercial generators working in continuous mode, for example, the Adelphi DDm neutron generator with integrated moderators~\cite{dd110m_and_dd109m}, which can produce thermal neutron flux up to $10^7$~n/cm$^2$/s.}{The interrogation source was a thermal neutron beam with a source strength of $10^6$~n/s, which can be achieved by the newest-generation neutron generators that can produce thermal neutron flux up to $10^7$~n/cm$^2$/s~\cite{dd110m_and_dd109m}}.

\deleted{Figure~18 shows the neutron singles and doubles as a function of remaining \textsuperscript{235}U mass in the partially spent pebble. Neutrons from the external source could be directly detected by the NMC, resulting in a constant background in the neutron singles. The number of induced fission is proportional to the amount of \textsuperscript{235}U in the fuel, resulting in a linear correlation between the neutron count rates and \textsuperscript{235}U mass.}
% \begin{figure}[!htbp]
%     \captionsetup{font=footnotesize}
%     \begin{subfigure}[t]{0.5\linewidth}
%         \centering
%         \includegraphics[width=\linewidth]{figs/active_singles.pdf}
%         \caption{Singles}
%         \label{fig:active_single}
%     \end{subfigure}\hfil
%     \begin{subfigure}[t]{0.5\linewidth}
%         \centering
%         \includegraphics[width=\linewidth]{figs/active_doubles.pdf}
%         \caption{Doubles}
%         \label{fig:active_double}
%     \end{subfigure}
%     \caption{\replaced{Simulated neutron}{Neutron} singles and doubles rate as a function of \textsuperscript{235}U mass in active interrogation mode.}
%     \label{fig:active_interrogation_results}
%     % \vspace{-1em}
% \end{figure}

% \subsection{\replaced{Passive Measurement}{Passive Interrogation}}
The \replaced{simulation setup of passive measurement}{passive interrogation setup} is similar to the one shown in Fig.~\ref{fig:active_interrogation_setup} with the interrogating neutron source removed. Instead, passive gamma-ray source, delayed \& alpha neutron source, and SF neutron source from the pebble itself were simulated. The spectrum and intensity of these sources were calculated based on the SCALE simulation in Section~\ref{sec:scale_simulation}.

\added{For each simulation, a list of neutron pulse timestamps was extracted from the MCNP output file~\cite{FANG2023109794}. Pulses arriving within \SI{180}{\nano\second} (measured dead-time of one channel) after the previous one in the same channel were rejected to account for the dead-time loss. We applied Eqs.~\eqref{eq:S_D_T} and \eqref{eq:S_D_uncertainty} to compute the neutron singles and doubles count rates due to each source term. The pre-delay time $T_{pd}$ was set to \SI{2}{\micro\second}, after which the Rossi-alpha distribution exhibited a decreasing trend; the gate width $G$ was set to \SI{48}{\micro\second}, which corresponded to a gate utilization factor of 0.81; the long delay time $T_{ld}$ was set to \SI{2}{\ms}.}

\begin{figure}[!htbp]
    \captionsetup{font=footnotesize}
    \begin{subfigure}[t]{0.5\linewidth}
        \centering
        \includegraphics[width=\linewidth]{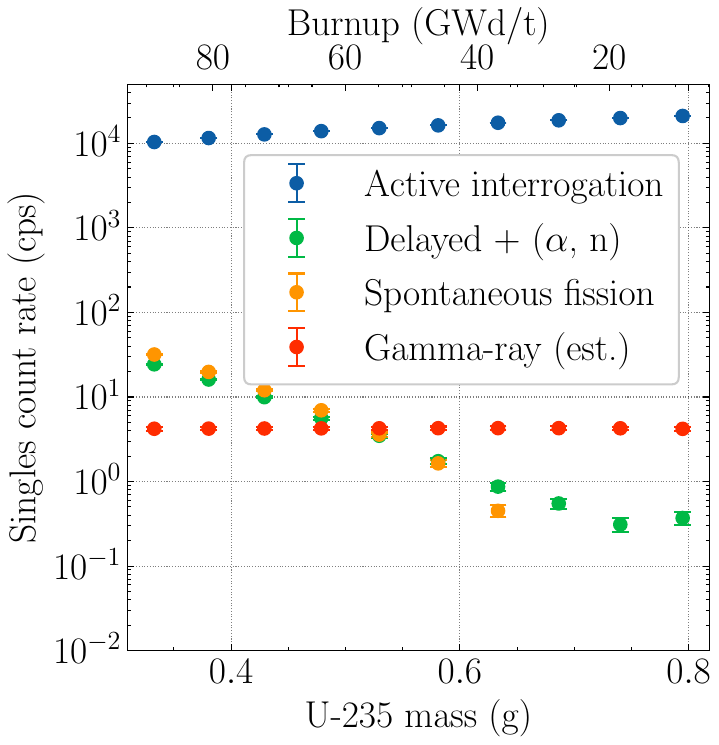}
        \caption{Singles}
        \label{fig:singles_comparison}
    \end{subfigure}\hfil
    \begin{subfigure}[t]{0.5\linewidth}
        \centering
        \includegraphics[width=\linewidth]{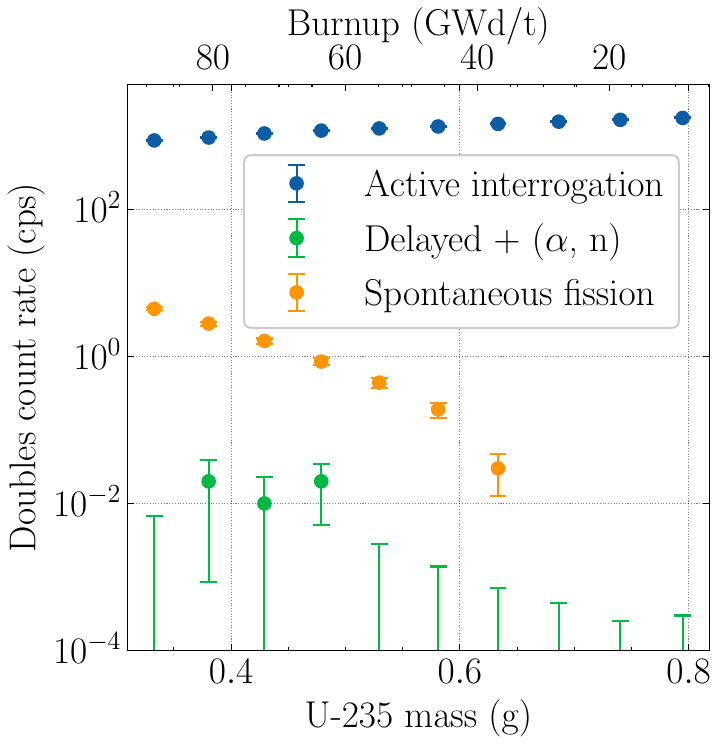}
        \caption{Doubles}
        \label{fig:doubles_comparison}
    \end{subfigure}
    \caption{Comparison of \added{simulated }neutron singles and doubles rates from various source terms as a function of \textsuperscript{235}U mass.\added{ Some green data points have y-values smaller than zero due to statistical fluctuation and are therefore invisible.}}
    \label{fig:comparison}
    % \vspace{-1em}
\end{figure}

\added{Figure~\ref{fig:comparison} shows the comparison of singles and doubles count rates from all source terms as a function of remaining \textsuperscript{235}U mass in the pebble with burnup ranging from 9 to \SI{90}{\gwdpert}. The singles and doubles due to active interrogation is shown as the blue dots. Neutrons from the external source can be directly detected by the NMC, resulting in a constant background in the neutron singles of approximately 2700~cps; on the other hand, background doubles rate due to the external beam is zero because the source emits one neutron at a time. \textsuperscript{235}U and \textsuperscript{239}Pu contribute over 99\% of the induced fission neutrons. The contribution of \textsuperscript{239}Pu increased from 1\% to 13\%, while the contribution of \textsuperscript{235}U decreased from 99\% to 86\%, when the fuel burnup increased from 9 to \SI{90}{\gwdpert}.}

\replaced{The singles contributed by gamma rays shown in Figure~\ref{fig:comparison}}{Figure~19 shows the singles count rate contributed by gamma rays, which} was obtained by multiplying the shielded gamma-ray intensity with the intrinsic gamma-ray efficiency of the system. The efficiency was assumed to be 10\textsuperscript{-12} based on the gamma-ray-insensitivity measurement results in Section~\ref{sec:gamma_insensitivity}. \deleted{Compared to count rates due to active interrogation, the gamma-ray count rate is negligible.}
% \begin{figure}[!htbp]
% 	\centering
% 	\includegraphics[width=.5\linewidth]{figs/singles_gamma-ray.pdf}
% 	\caption{\replaced{Simulated singles}{Singles} count rate \replaced{resulting}{resulted} from gamma-rays.}
% 	\label{fig:singles_gamma}
% \end{figure}

\deleted{Figure~20 shows the the singles and doubles contributed by delayed \& {($\alpha$, n)} neutrons and Fig.~21 shows the singles and doubles contributed by spontaneous fissioning nuclides. }\replaced{The singles count rates contributed by the delayed \& {($\alpha$,n)} neutrons and spontaneous fissioning nuclides both increase}{Both singles count rates increase} with the depletion of uranium due to the accumulation of fission products. The delayed \& \replaced{($\alpha$,n)}{($\alpha$, n)} neutrons are not correlated in time, therefore the resulting doubles count rate comes only from \replaced{statistical fluctuation}{random coincidence} and is always near zero\deleted{ with large fluctuation, as shown in Fig.~20}. The doubles count rate \replaced{resulting}{resulted} from spontaneous fission decreases with \textsuperscript{235}U mass because the fraction of SF nuclides increase as the fuel burns.\deleted{ Nevertheless, these count rates are negligible compared to the count rates contributed by the external interrogation source shown in Fig.~18}.
% \begin{figure}[!htbp]
%     \captionsetup{font=footnotesize}
%     \begin{subfigure}[t]{0.47\linewidth}
%         \centering
%         \includegraphics[width=\linewidth]{figs/singles_delayed_alpha.pdf}
%         \caption{Singles}
%         \label{fig:singles_delayed_alpha}
%     \end{subfigure}\hfil
%     \begin{subfigure}[t]{0.5\linewidth}
%         \centering
%         \includegraphics[width=\linewidth]{figs/doubles_delayed_alpha.pdf}
%         \caption{Doubles}
%         \label{fig:doubles_delayed_alpha}
%     \end{subfigure}
%     \caption{\replaced{Simulated neutron}{Neutron} singles and doubles rate \replaced{resulting}{resulted} from delayed and \replaced{($\alpha$,n)}{($\alpha$, n)} neutrons as a function of \textsuperscript{235}U mass.}
%     \label{fig:delayed_alpha_results}
%     % \vspace{-1em}
% \end{figure}

% \begin{figure}[!htbp]
%     \captionsetup{font=footnotesize}
%     \begin{subfigure}[t]{0.5\linewidth}
%         \centering
%         \includegraphics[width=\linewidth]{figs/singles_sf.pdf}
%         \caption{Singles}
%         \label{fig:singles_sf}
%     \end{subfigure}\hfil
%     \begin{subfigure}[t]{0.5\linewidth}
%         \centering
%         \includegraphics[width=\linewidth]{figs/doubles_sf.pdf}
%         \caption{Doubles}
%         \label{fig:doubles_sf}
%     \end{subfigure}
%     \caption{\replaced{Simulated neutron}{Neutron} singles and doubles rate \replaced{resulting}{resulted} from SF neutrons as a function of \textsuperscript{235}U mass.}
%     \label{fig:sf_results}
%     % \vspace{-1em}
% \end{figure}

\added{Overall, for pebbles with burnup below \SI{90}{\gwdpert}, singles and doubles from active interrogation constitute more than 99\% and 99.5\% of the total, respectively. Therefore, the impact of gamma-ray background and passive neutrons is negligible and our system is feasible for assay of \textsuperscript{235}U in used TRISO-fueled pebbles.}
\subsection{\replaced{Assay error and uncertainty}{Comparison}}
\deleted{Figure~\ref{fig:comparison} shows the comparison of singles and doubles count rate from all source terms. For pebbles with burnup below 90~GWd/t, singles and doubles from active interrogation constitute more than 99\% and 99.5\% of the total, respectively. Therefore, the impact of gamma-ray background and passive neutrons is negligible and our system is feasible for assay of \textsuperscript{235}U in used TRISO-fueled pebbles.}

\begin{figure}[!htbp]
    \captionsetup{font=footnotesize}
    \begin{subfigure}[t]{0.5\linewidth}
        \centering
        \includegraphics[width=\linewidth]{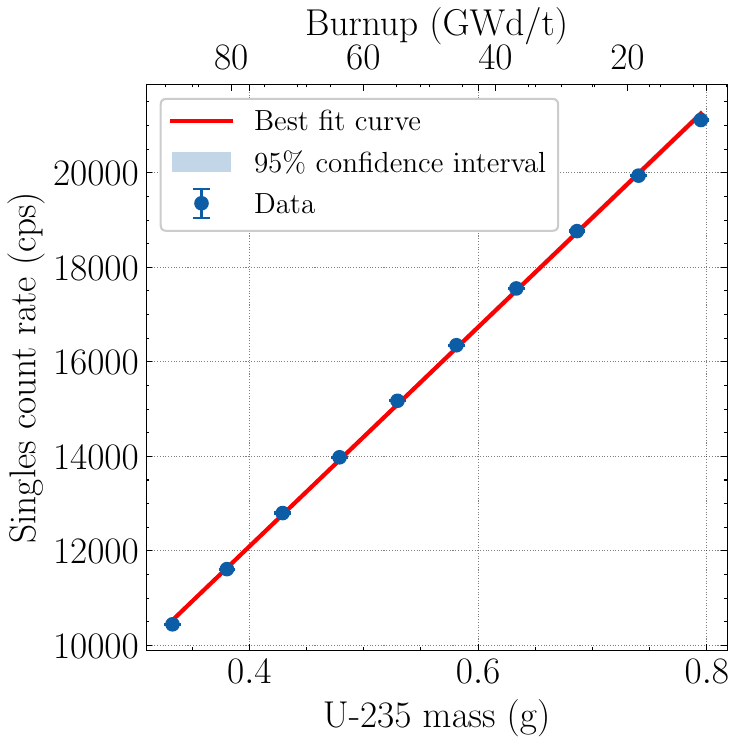}
        \caption{Singles}
        \label{fig:single_total}
    \end{subfigure}\hfil
    \begin{subfigure}[t]{0.5\linewidth}
        \centering
        \includegraphics[width=\linewidth]{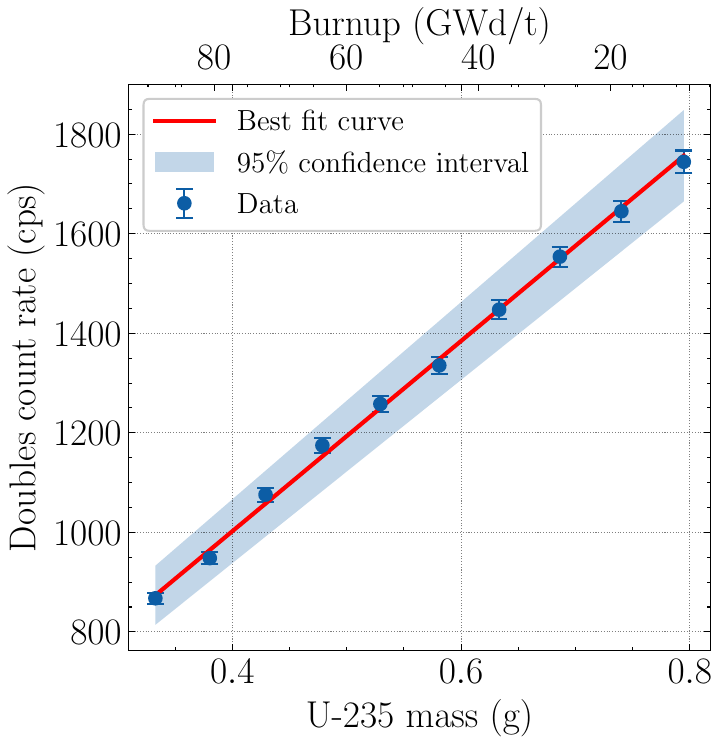}
        \caption{Doubles}
        \label{fig:double_total}
    \end{subfigure}
    \caption{{Simulated neutron} singles and doubles rate from all source terms as a function of \textsuperscript{235}U mass in the spent fuel pebble.}
    \label{fig:linear_fits}
    % \vspace{-1em}
\end{figure}

\added{For each burnup step, we merged the pulse timestamp lists from different source terms into a single list and calculated the total singles and doubles rates by applying the shift-register algorithm. }As shown in Fig.~\ref{fig:linear_fits}, a calibration curve is obtained by fitting a linear function to the data. The \textsuperscript{235}U mass in an unknown fuel sample can be estimated by finding the point on the calibration curve corresponding to the doubles count rate measured by the NMC~\cite{FANG2023109794}. 

\begin{figure}[!htbp]
    \captionsetup{font=footnotesize}
    \begin{subfigure}[t]{0.5\linewidth}
        \centering
        \includegraphics[width=\linewidth]{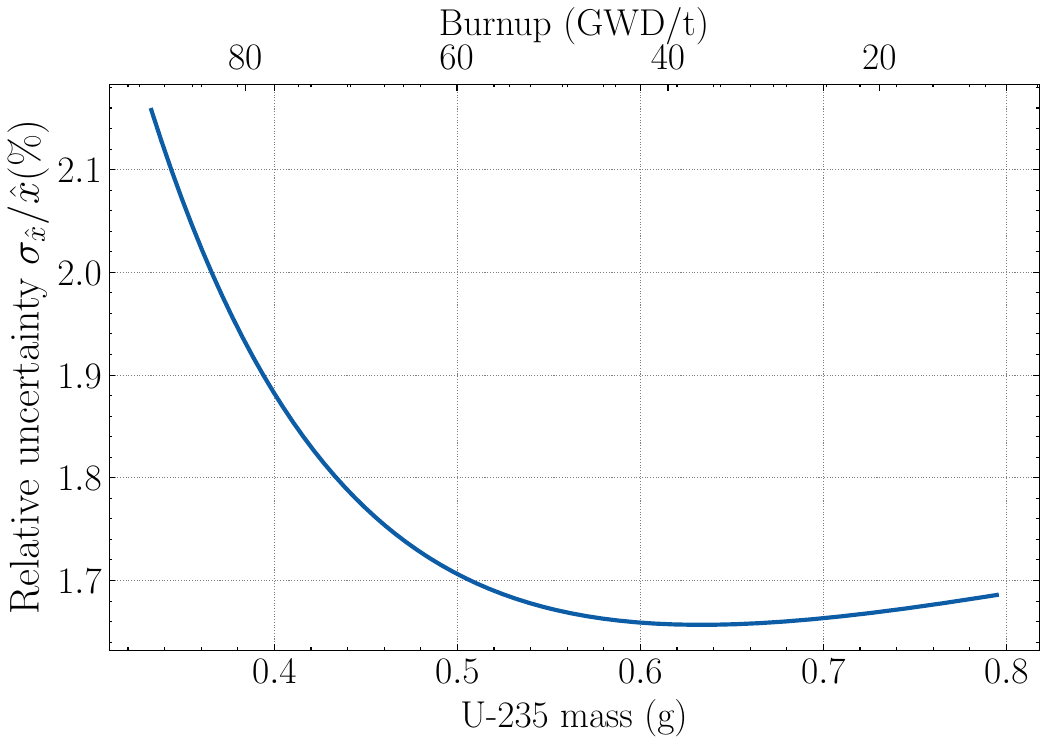}
        \caption{Uncertainty}
        \label{fig:relative_uncertainty}
    \end{subfigure}\hfil
    \begin{subfigure}[t]{0.5\linewidth}
        \centering
        \includegraphics[width=\linewidth]{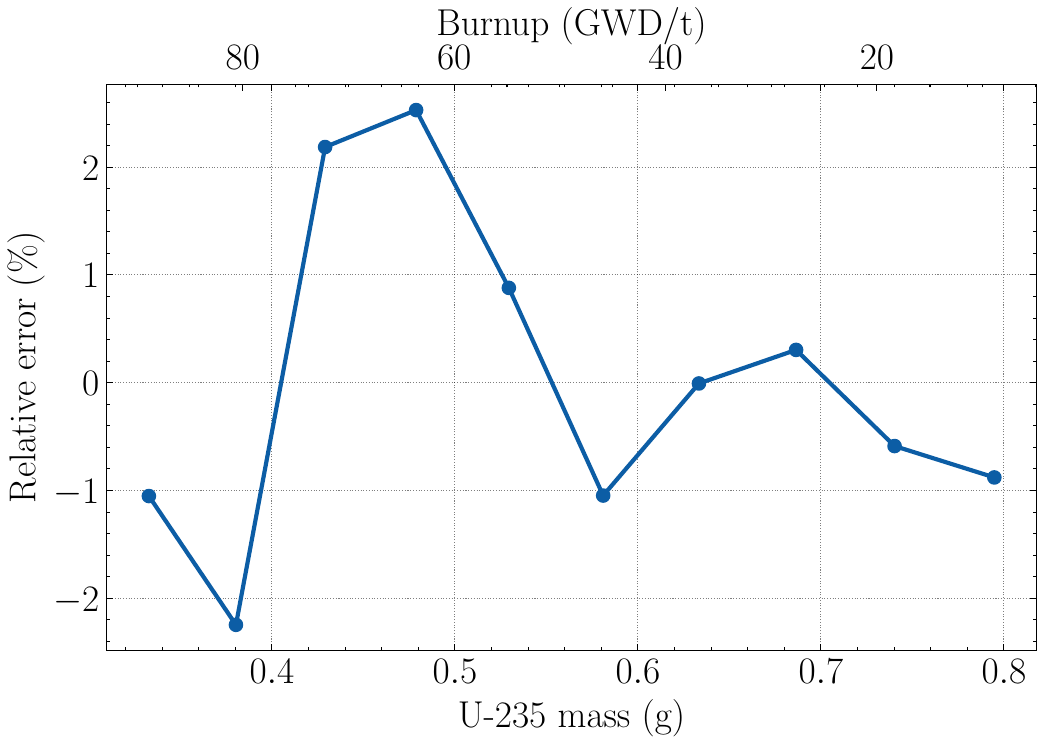}
        \caption{Error}
        \label{fig:relative_error}
    \end{subfigure}
    \caption{Comparison of relative uncertainty and relative error associated with the \textsuperscript{235}U mass assay. The assay time was \replaced{\SI{100}{\s}}{varied from 100~s to 500~s in step of 100~s}.}
    \label{fig:assay_quality}
    % \vspace{-1em}
\end{figure}

Figure~\ref{fig:relative_uncertainty} shows the relative uncertainty of the \textsuperscript{235}U mass estimation, obtained by propagating the uncertainty of the calibration curve and uncertainty of measured doubles. Figure~\ref{fig:relative_error} shows the relative error associated with \textsuperscript{235}U mass estimation, which is defined as the relative difference between the estimated mass and true mass. Both the relative uncertainty and relative error associated with the \textsuperscript{235}U mass assay are below 2.5\% with an inspection time of \SI{100}{\s}. 

\section{Discussions and Conclusions}\label{sec:conclusion}
In this work, we have developed a high-fidelity Monte Carlo model of the HTR-10 reactor in SCALE~6.2.4 and performed fuel burnup calculation. Fuel composition, passive gamma-ray and neutron source terms were extracted for used fuel pebbles with burnup between \SI{8.88}{\gwdpert} and \SI{88.88}{\gwdpert}. Simulation showed that an intrinsic gamma-ray efficiency of 10\textsuperscript{-12} is required of the NMC to prevent the passive gamma-ray background from interfering with the neutron measurements. We have experimentally demonstrated that a gamma-ray-efficiency of 10\textsuperscript{-12} can be achieved by our BCS-based NMC, with a threshold of \SI{32}{\mV}, under a high gamma-ray exposure rate of \SI[per-mode = symbol]{340.87}{\Roentgen\per\hour}\added{ (\SI[per-mode = symbol]{8.3e-4}{\gray\per\s})}, comparable to the exposure rate that would be caused by used TRISO fueled pebbles. We simulated the neutron interrogation of used fuel pebbles and found that the contributions to the singles and doubles by passive neutron and gamma-ray emitters were negligible compared to induced fission.  Therefore, measurement of \textsuperscript{235}U mass in used TRISO-fueled pebble using our BCS-NMC is feasible. The relative uncertainty and error associated with the \textsuperscript{235}U mass assay is below 2.5\% for pebbles with burnup below \SI{90}{\gwdpert} with \replaced{a measurement time}{an interrogation time} of \SI{100}{\s}, which is compatible with practical reactor operation.

\section*{ACKNOWLEDGMENTS}
This work was funded in part by STTR-DOE grant DE-SC0020733. We would like to thank Dr. Joe Love and Carle Foundation Hospital for the assistance in the measurement of \textsuperscript{192}Ir source.

\bibliography{references}

\begin{thebibliography}{10}
\expandafter\ifx\csname url\endcsname\relax
  \def\url#1{\texttt{#1}}\fi
\expandafter\ifx\csname urlprefix\endcsname\relax\def\urlprefix{URL }\fi
\expandafter\ifx\csname href\endcsname\relax
  \def\href#1#2{#2} \def\path#1{#1}\fi

\bibitem{MOORE1982153}
R.~Moore, M.~Kantor, H.~Brey, H.~Olson,
  \href{https://www.sciencedirect.com/science/article/pii/0029549382902126}{Htgr
  experience, programs, and future applications}, Nuclear Engineering and
  Design 72~(2) (1982) 153--174.
\newblock \href {https://doi.org/https://doi.org/10.1016/0029-5493(82)90212-6}
  {\path{doi:https://doi.org/10.1016/0029-5493(82)90212-6}}.
\newline\urlprefix\url{https://www.sciencedirect.com/science/article/pii/0029549382902126}

\bibitem{international2010iaea}
\href{https://www.iaea.org/publications/8270/high-temperature-gas-cooled-reactor-fuels-and-materials}{High
  Temperature Gas Cooled Reactor Fuels and Materials}, no. 1645 in TECDOC
  Series (CD-ROM), INTERNATIONAL ATOMIC ENERGY AGENCY, Vienna, 2010.
\newline\urlprefix\url{https://www.iaea.org/publications/8270/high-temperature-gas-cooled-reactor-fuels-and-materials}

\bibitem{JASZCZUR20167861}
M.~Jaszczur, M.~A. Rosen, T.~Śliwa, M.~Dudek, L.~Pieńkowski,
  \href{https://www.sciencedirect.com/science/article/pii/S0360319916308710}{Hydrogen
  production using high temperature nuclear reactors: Efficiency analysis of a
  combined cycle}, International Journal of Hydrogen Energy 41~(19) (2016)
  7861--7871, special Issue on Progress in Hydrogen Production and Applications
  (ICH2P-2015), 3-6 May 2015, Oshawa, Ontario, Canada.
\newblock \href
  {https://doi.org/https://doi.org/10.1016/j.ijhydene.2015.11.190}
  {\path{doi:https://doi.org/10.1016/j.ijhydene.2015.11.190}}.
\newline\urlprefix\url{https://www.sciencedirect.com/science/article/pii/S0360319916308710}

\bibitem{DEMKOWICZ2019434}
P.~A. Demkowicz, B.~Liu, J.~D. Hunn,
  \href{https://www.sciencedirect.com/science/article/pii/S0022311518310213}{Coated
  particle fuel: Historical perspectives and current progress}, Journal of
  Nuclear Materials 515 (2019) 434--450.
\newblock \href {https://doi.org/https://doi.org/10.1016/j.jnucmat.2018.09.044}
  {\path{doi:https://doi.org/10.1016/j.jnucmat.2018.09.044}}.
\newline\urlprefix\url{https://www.sciencedirect.com/science/article/pii/S0022311518310213}

\bibitem{KANIA2013545}
M.~J. Kania, H.~Nabielek, K.~Verfondern, H.-J. Allelein,
  \href{https://www.sciencedirect.com/science/article/pii/S0022311513008088}{Testing
  of htr uo2 triso fuels in avr and in material test reactors}, Journal of
  Nuclear Materials 441~(1) (2013) 545--562.
\newblock \href {https://doi.org/https://doi.org/10.1016/j.jnucmat.2013.05.062}
  {\path{doi:https://doi.org/10.1016/j.jnucmat.2013.05.062}}.
\newline\urlprefix\url{https://www.sciencedirect.com/science/article/pii/S0022311513008088}

\bibitem{kadak2005future}
A.~C. Kadak, A future for nuclear energy: pebble bed reactors, International
  journal of critical infrastructures 1~(4) (2005) 330--345.
\newblock \href {https://doi.org/10.1016/j.pnucene.2022.104175}
  {\path{doi:10.1016/j.pnucene.2022.104175}}.

\bibitem{whetstone2014review}
Z.~Whetstone, K.~Kearfott, A review of conventional explosives detection using
  active neutron interrogation, Journal of Radioanalytical and Nuclear
  Chemistry 301 (2014) 629--639.

\bibitem{chichester2008using}
D.~L. Chichester, E.~H. Seabury, Using electronic neutron generators in active
  interrogation to detect shielded fissionable material, in: 2008 IEEE Nuclear
  Science Symposium Conference Record, IEEE, 2008, pp. 3361--3367.

\bibitem{myers2005photon}
W.~L. Myers, C.~A. Goulding, C.~L. Hollas, C.~E. Moss, Photon and neutron
  active interrogation of highly enriched uranium, in: AIP Conference
  Proceedings, Vol. 769, American Institute of Physics, 2005, pp. 1688--1692.

\bibitem{gozani1981active}
T.~Gozani, Active nondestructive assay of nuclear materials: principles and
  applications, Tech. rep., Science Applications, Inc., Palo Alto, CA (USA);
  Monsanto Research Corp~… (1981).

\bibitem{melton2000calibration}
S.~Melton, Calibration of the crated waste assay monitor (cwam) for the
  low-level waste measurements for the y-12 plant, Tech. rep., Los Alamos
  National Lab.(LANL), Los Alamos, NM (United States) (2000).

\bibitem{eccleston1979measurement}
G.~Eccleston, H.~Menlove, M.~Echo, Measurement system for high enriched spent
  fuel assemblies and waste solids, Tech. rep., Los Alamos National Lab.(LANL),
  Los Alamos, NM (United States) (1979).

\bibitem{reilly1991passive}
D.~Reilly, N.~Ensslin, H.~Smith~Jr, S.~Kreiner, Passive nondestructive assay of
  nuclear materials, Tech. rep., Nuclear Regulatory Commission (1991).

\bibitem{DIFULVIO201792}
A.~{Di Fulvio}, T.~Shin, T.~Jordan, C.~Sosa, M.~Ruch, S.~Clarke, D.~Chichester,
  S.~Pozzi,
  \href{https://www.sciencedirect.com/science/article/pii/S0168900217303066}{Passive
  assay of plutonium metal plates using a fast-neutron multiplicity counter},
  Nuclear Instruments and Methods in Physics Research Section A: Accelerators,
  Spectrometers, Detectors and Associated Equipment 855 (2017) 92--101.
\newblock \href {https://doi.org/https://doi.org/10.1016/j.nima.2017.02.082}
  {\path{doi:https://doi.org/10.1016/j.nima.2017.02.082}}.
\newline\urlprefix\url{https://www.sciencedirect.com/science/article/pii/S0168900217303066}

\bibitem{FANG2023109794}
M.~Fang, J.~Lacy, A.~Athanasiades, A.~{Di Fulvio},
  \href{https://www.sciencedirect.com/science/article/pii/S0306454923001135}{Boron
  coated straw-based neutron multiplicity counter for neutron interrogation of
  triso fueled pebbles}, Annals of Nuclear Energy 187 (2023) 109794.
\newblock \href {https://doi.org/https://doi.org/10.1016/j.anucene.2023.109794}
  {\path{doi:https://doi.org/10.1016/j.anucene.2023.109794}}.
\newline\urlprefix\url{https://www.sciencedirect.com/science/article/pii/S0306454923001135}

\bibitem{topan2016study}
S.~Topan, I.~Dwi, et~al., Study on fuel multipass effect on core performance of
  small pebble bed reactor (2016).

\bibitem{su2004design}
B.~Su, Design and construction of a prototype advanced on-line fuel burn-up
  monitoring system for the modular pebble bed reactor, Tech. rep., University
  of Cincinnati, Cincinnati, OH (US) (2004).

\bibitem{rearden2018scale}
B.~T. Rearden, M.~A. Jessee, Scale code system, Tech. rep., Oak Ridge National
  Lab.(ORNL), Oak Ridge, TN (United States) (2018).

\bibitem{xu2002htr}
Y.~Xu, The htr-10 project and its further development, Tech. rep. (2002).

\bibitem{terry2005evaluation}
W.~K. Terry, Evaluation of the initial critical configuration of the htr-10
  pebble-bed reactor, Tech. rep., Idaho National Lab.(INL), Idaho Falls, ID
  (United States) (2005).

\bibitem{demkowicz2019triso}
P.~A. Demkowicz, Triso fuel: design, manufacturing, and performance, Tech.
  rep., Idaho National Lab.(INL), Idaho Falls, ID (United States) (2019).

\bibitem{barrachin2010high}
M.~Barrachin, V.~Basini, {\"U}.~C. BU, R.~Dubourg, M.~Feltus, D.~Greneche,
  P.~Guillermier, U.~Hansen, D.~Hanson, J.~Hunn, et~al., High temperature gas
  cooled reactor fuels and materials, IAEATECDOC-1645 (2010).

\bibitem{colak2005monte}
{\"U}.~Colak, V.~Seker, Monte carlo criticality calculations for a pebble bed
  reactor with mcnp, Nuclear science and engineering 149~(2) (2005) 131--137.

\bibitem{international2004iaea}
\href{https://www.iaea.org/publications/6821/evaluation-of-high-temperature-gas-cooled-reactor-performance-benchmark-analysis-related-to-initial-testing-of-the-httr-and-htr-10}{Evaluation
  of High Temperature Gas Cooled Reactor Performance: Benchmark Analysis
  Related to Initial Testing of the HTTR and HTR-10}, no. 1382 in TECDOC
  Series, INTERNATIONAL ATOMIC ENERGY AGENCY, Vienna, 2004.
\newline\urlprefix\url{https://www.iaea.org/publications/6821/evaluation-of-high-temperature-gas-cooled-reactor-performance-benchmark-analysis-related-to-initial-testing-of-the-httr-and-htr-10}

\bibitem{sunny2010scale}
E.~E. Sunny, G.~Ilas, Scale 6 analysis of htr-10 pebble-bed reactor for initial
  critical configuration, in: Proc. of PHYSOR, 2010.

\bibitem{ilas2010scale}
G.~Ilas, On scale validation for pbr analysis, Tech. rep., Oak Ridge National
  Lab.(ORNL), Oak Ridge, TN (United States) (2010).

\bibitem{yang2002fuel}
Y.~Yang, Z.~Luo, X.~Jing, Z.~Wu, Fuel management of the htr-10 including the
  equilibrium state and the running-in phase, Nuclear Engineering and Design
  218~(1-3) (2002) 33--41.

\bibitem{SU2006686}
B.~Su, Z.~Zhao, J.~Chen, A.~I. Hawari,
  \href{https://www.sciencedirect.com/science/article/pii/S0149197006000771}{Assessment
  of on-line burnup monitoring of pebble bed reactor fuel by passive neutron
  counting}, Progress in Nuclear Energy 48~(7) (2006) 686--702.
\newblock \href {https://doi.org/https://doi.org/10.1016/j.pnucene.2006.06.013}
  {\path{doi:https://doi.org/10.1016/j.pnucene.2006.06.013}}.
\newline\urlprefix\url{https://www.sciencedirect.com/science/article/pii/S0149197006000771}

\bibitem{CHEN2003393}
J.~Chen, A.~I. Hawari, Z.~Zhao, B.~Su,
  \href{https://www.sciencedirect.com/science/article/pii/S0168900203011057}{Gamma-ray
  spectrometry analysis of pebble bed reactor fuel using monte carlo
  simulations}, Nuclear Instruments and Methods in Physics Research Section A:
  Accelerators, Spectrometers, Detectors and Associated Equipment 505~(1)
  (2003) 393--396, proceedings of the tenth Symposium on Radiation Measurements
  and Applications.
\newblock \href {https://doi.org/https://doi.org/10.1016/S0168-9002(03)01105-7}
  {\path{doi:https://doi.org/10.1016/S0168-9002(03)01105-7}}.
\newline\urlprefix\url{https://www.sciencedirect.com/science/article/pii/S0168900203011057}

\bibitem{YAN2014172}
W.-H. Yan, L.-G. Zhang, Z.~Zhang, Y.~Zhang, Z.-G. Xiao,
  \href{https://www.sciencedirect.com/science/article/pii/S0029549313007176}{Prototype
  studies on the nondestructive online burnup determination for the modular
  pebble bed reactors}, Nuclear Engineering and Design 267 (2014) 172--179.
\newblock \href
  {https://doi.org/https://doi.org/10.1016/j.nucengdes.2013.12.036}
  {\path{doi:https://doi.org/10.1016/j.nucengdes.2013.12.036}}.
\newline\urlprefix\url{https://www.sciencedirect.com/science/article/pii/S0029549313007176}

\bibitem{VERGARI2021111189}
L.~Vergari, M.~Fratoni,
  \href{https://www.sciencedirect.com/science/article/pii/S0029549321001412}{Spent
  fuel management strategies for fluoride-cooled pebble bed reactors}, Nuclear
  Engineering and Design 378 (2021) 111189.
\newblock \href
  {https://doi.org/https://doi.org/10.1016/j.nucengdes.2021.111189}
  {\path{doi:https://doi.org/10.1016/j.nucengdes.2021.111189}}.
\newline\urlprefix\url{https://www.sciencedirect.com/science/article/pii/S0029549321001412}

\bibitem{LIANG20201}
J.~Liang, B.~Singh, E.~McCutchan, I.~Dillmann, M.~Birch, A.~Sonzogni, X.~Huang,
  M.~Kang, J.~Wang, G.~Mukherjee, K.~Banerjee, D.~Abriola, A.~Algora, A.~Chen,
  T.~Johnson, K.~Miernik,
  \href{https://www.sciencedirect.com/science/article/pii/S009037522030034X}{Compilation
  and evaluation of beta-delayed neutron emission probabilities and half-lives
  for z > 28 precursors}, Nuclear Data Sheets 168 (2020) 1--116.
\newblock \href {https://doi.org/https://doi.org/10.1016/j.nds.2020.09.001}
  {\path{doi:https://doi.org/10.1016/j.nds.2020.09.001}}.
\newline\urlprefix\url{https://www.sciencedirect.com/science/article/pii/S009037522030034X}

\bibitem{SHURSHIKOV1985509}
E.~Shurshikov, M.~Filchenkov, Y.~Jaborov, A.~Khovanovich,
  \href{https://www.sciencedirect.com/science/article/pii/S0090375285800272}{Nuclear
  data sheets for a = 242}, Nuclear Data Sheets 45~(3) (1985) 509--556.
\newblock \href {https://doi.org/https://doi.org/10.1016/S0090-3752(85)80027-2}
  {\path{doi:https://doi.org/10.1016/S0090-3752(85)80027-2}}.
\newline\urlprefix\url{https://www.sciencedirect.com/science/article/pii/S0090375285800272}

\bibitem{cristallo2018importance}
S.~Cristallo, M.~La~Cognata, C.~Massimi, A.~Best, S.~Palmerini, O.~Straniero,
  O.~Trippella, M.~Busso, G.~Ciani, F.~Mingrone, et~al., The importance of the
  13c ($\alpha$, n) 16o reaction in asymptotic giant branch stars, The
  astrophysical journal 859~(2) (2018) 105.

\bibitem{langner1998application}
D.~Langner, J.~Stewart, M.~Pickrell, M.~Krick, N.~Ensslin, W.~Harker,
  Application guide to neutron multiplicity counting, Tech. rep., Los Alamos
  National Laboratory, Los Alamos, NM (1998).

\bibitem{Croft2012152}
S.~Croft, D.~Henzlova, D.~Hauck, Extraction of correlated count rates using
  various gate generation techniques: Part i theory, NIMA 691 (2012) 152 –
  158.
\newblock \href {https://doi.org/10.1016/j.nima.2012.06.011}
  {\path{doi:10.1016/j.nima.2012.06.011}}.

\bibitem{santi2014china}
P.~A. Santi, K.~A. Miller, P.~J. Karpius, D.~T. Vo, D.~J. Mercer, K.~C. Frame,
  A.~P. Belian, H.~A. Nordquist, M.~T. Swinhoe, China center of excellence
  nondestructive assay training course, Tech. rep., Los Alamos National
  Lab.(LANL), Los Alamos, NM (United States) (2014).

\bibitem{prasad2018analytical}
M.~Prasad, N.~Snyderman, J.~Verbeke, Analytical error bars and rsd for neutron
  multiplicity counting, Nuclear Instruments and Methods in Physics Research
  Section A: Accelerators, Spectrometers, Detectors and Associated Equipment
  903 (2018) 25--31.
\newblock \href {https://doi.org/https://doi.org/10.1016/j.nima.2018.06.010}
  {\path{doi:https://doi.org/10.1016/j.nima.2018.06.010}}.

\bibitem{knoll2010radiation}
G.~F. Knoll, Radiation detection and measurement, John Wiley \& Sons, 2010.

\bibitem{afterloader}
Elekta flexitron afterloader,
  \url{https://www.elekta.com/products/brachytherapy/flexitron/}, accessed:
  2023-03-27.

\bibitem{lopes2022vivo}
A.~Lopes, E.~Sabondjian, A.~R. Baltazar, In vivo dosimetry for superficial high
  dose rate brachytherapy with optically stimulated luminescence dosimeters: A
  comparison study with metal-oxide-semiconductor field-effect transistors,
  Radiation 2~(4) (2022) 338--356.

\bibitem{KOUZES2011412}
R.~T. Kouzes, J.~H. Ely, A.~T. Lintereur, E.~K. Mace, D.~L. Stephens, M.~L.
  Woodring,
  \href{https://www.sciencedirect.com/science/article/pii/S0168900211014628}{Neutron
  detection gamma ray sensitivity criteria}, Nuclear Instruments and Methods in
  Physics Research Section A: Accelerators, Spectrometers, Detectors and
  Associated Equipment 654~(1) (2011) 412--416.
\newblock \href {https://doi.org/https://doi.org/10.1016/j.nima.2011.07.030}
  {\path{doi:https://doi.org/10.1016/j.nima.2011.07.030}}.
\newline\urlprefix\url{https://www.sciencedirect.com/science/article/pii/S0168900211014628}

\bibitem{AKhaplanov_2013}
A.~Khaplanov, F.~Piscitelli, J.~C. Buffet, J.~F. Clergeau, J.~Correa, P.~van
  Esch, M.~Ferraton, B.~Guerard, R.~Hall-Wilton,
  \href{https://dx.doi.org/10.1088/1748-0221/8/10/P10025}{Investigation of
  gamma-ray sensitivity of neutron detectors based on thin converter films},
  Journal of Instrumentation 8~(10) (2013) P10025.
\newblock \href {https://doi.org/10.1088/1748-0221/8/10/P10025}
  {\path{doi:10.1088/1748-0221/8/10/P10025}}.
\newline\urlprefix\url{https://dx.doi.org/10.1088/1748-0221/8/10/P10025}

\bibitem{osti_1419730}
C.~J. Werner, J.~S. Bull, C.~J. Solomon, F.~B. Brown, G.~W. McKinney, M.~E.
  Rising, D.~A. Dixon, R.~L. Martz, H.~G. Hughes, L.~J. Cox, A.~J. Zukaitis,
  J.~C. Armstrong, R.~A. Forster, L.~Casswell,
  \href{https://www.osti.gov/biblio/1419730}{Mcnp version 6.2 release notes} (2
  2018).
\newblock \href {https://doi.org/10.2172/1419730} {\path{doi:10.2172/1419730}}.
\newline\urlprefix\url{https://www.osti.gov/biblio/1419730}

\bibitem{dd110m_and_dd109m}
A.~Technology, Dd110m and dd109m,
  \url{https://adelphitech.com/products/deuterium-deuterium-dd-neutron-generators/integrated-moderators-dd110-m-dd-109-m/},
  accessed: 2023-06-30.

\end{thebibliography}

\end{document}